\documentclass[apj]{emulateapj}
\usepackage{graphicx}
\usepackage{xfrac}
\usepackage{amsmath}    
\usepackage{amssymb} 

\def\teff {{$T_\mathrm{eff}$ }}

\def\Re {{$\,R_\oplus$ }}
\def\Me {{$\,M_\oplus$ }}

\def\Mj {{$\,M_\mathrm{J}$ }}
\def\lg {{log$\,g$ }}
\def\pl {{planetary }}
\def\plr {{planetary radius }}

\def \hsm {{host star metallicity }}
\def \hs {{host star }}
\def \sc {{SWEET-Cat }}

\begin{document}

\title{Properties and Occurrence Rates of  $Kepler$ Exoplanet Candidates as a Function of Host Star Metallicity from the DR25 Catalog}
\shorttitle{ Occurrence rate of planets as a function of Host star metallicity }

\author{Mayank Narang\altaffilmark{1}}
\affil{Department of Astronomy and Astrophysics, Tata Institute of Fundamental Research \\
Homi Bhabha Road, Colaba, Mumbai 400005, India}

\author{ P. Manoj}
\affil{Department of Astronomy and Astrophysics, Tata Institute of Fundamental Research \\
Homi Bhabha Road, Colaba, Mumbai 400005, India}

\author{E. Furlan}
\affil{IPAC, Mail Code 314-6, Caltech, 1200 E. California Blvd., Pasadena, CA 91125, USA}

\author{ C. Mordasini}
\affil{Physikalisches Institut, Universit\"{a}t Bern, Gesellschaftstrasse 6, 3012 Bern, Switzerland}

\author{ Thomas Henning\altaffilmark{2} }
\affil{Max-Planck-Institut f\"{u}r Astronomie, K\"{o}nigstuhl 17, D-69117 Heidelberg, Germany\\
}

\author{ Blesson Mathew}
\affil{Department of Physics, Christ University, Hosur Road, Bangalore 560029, India}

\author{Ravinder K. Banyal}
\affil{Indian Institute of Astrophysics, Bangalore, 560034 India}

\author{T. SIVARANI }
\affil{Indian Institute of Astrophysics, Bangalore, 560034 India}

\shortauthors{Mayank. N et al.}
\altaffiltext{1}{mayank.narang@tifr.res.in}
\altaffiltext{2}{Adjunct Professor, Department of Astronomy and Astrophysics, Tata Institute of Fundamental Research 
Homi Bhabha Road, Colaba, Mumbai 400005, India}

\begin{abstract}
Correlations between the occurrence rate of exoplanets and their host star properties provide important clues about the planet formation processes. We studied the dependence of the observed properties of exoplanets (radius, mass, and orbital period) as a function of their host star metallicity. We analyzed the planetary radii and orbital periods of over 2800 $Kepler$ candidates from the latest $Kepler$ data release DR25 (Q1-Q17) with revised planetary radii based on $Gaia$~DR2 as a function of host star metallicity (from the Q1-Q17 (DR25) stellar and planet catalog). With a much larger sample and improved radius measurements, we are able to reconfirm previous results in the literature. We show that the average metallicity of the host star increases as the radius of the planet increases. We demonstrate this by first calculating the average host star metallicity for different radius bins and then supplementing these results by calculating the occurrence rate as a function of planetary radius and host star metallicity.  We find a similar trend between host star metallicity and planet mass: the average host star metallicity increases with increasing planet mass. This trend, however, reverses for masses $> 4.0\, M_\mathrm{J}$: host star metallicity drops with increasing planetary mass. We further examined the correlation between the host star metallicity and the orbital period of the planet. We find that for planets with orbital periods less than 10 days, the average metallicity of the host star is higher than that for planets with periods greater than 10 days.
\end{abstract}


\section{Introduction } 
\label{sec intro}

The launch of the $Kepler$ Mission has provided us with an unprecedented view of planetary systems around stars other than the Sun. $Kepler$ has added significantly to the number of smaller (\pl radius, $R_\mathrm{P}$ $\leq$ 4$\,R_\oplus$) planets known to date  \citep[e.g.,][]{Borucki10,batalha11,Lissauer11,borucki12,howard12,batalha13,Fabrycky14}. The detection of more than 4000 $Kepler$ planet candidates has ushered in an era of statistically significant studies of exoplanet properties \citep[e.g.,][]{howard12,dressing13,petigura13,mann13,morton14,mulders15drop,mulders15small,burke15,christiansen15,wang15,CKS3,CKS4,pascucci18}.

Studies of exoplanet properties as a function of host star properties are particularly interesting as they provide important clues to the formation of planetary systems. Planetary systems are formed out of protoplanetary disks surrounding young stars. These disks are the byproducts of the star formation process and both the star and the disk (and the planetary system) are formed out of the same molecular cloud. The disk properties are known to strongly correlate with the host star properties \citep[e.g.,][]{muz2010,mcclure10,andrews11,manoj11,furlan11,Andrews13,kim13,mohanty13,kim16,pascucci16}.

Observed properties of exoplanets are also correlated with the host star properties. For example, the occurrence rate of exoplanets has a strong dependence on the spectral type of the host stars: the occurrence rate of giant planets $(R_\mathrm{P} > 4-6\,R_\oplus)$ is found to be higher for F, G, and K stars compared to that for M type stars \citep{cumming08,howard12,bonfils13,gaidos13,mulders15small,winn15,Obermeier16,mulders18,winn18}.

Radial velocity studies prior to $Kepler$ have demonstrated a strong correlation between the occurrence rate of giant planets ($M_P > 0.5\, M_\mathrm{J}$) and the metallicity of their host stars \citep{gonzalez97,santos01,santos04,fischer05,johnson07, Sousa08,sozzetti09,johnson10,ghezzi10,sousa11,mortier12,mulders18,winn18}. These studies found that as the metallicity of the host stars increases the fraction/frequency of giant planets around them also increases. Further, \cite{fischer05} and \cite{johnson10} showed that the fraction of stars hosting gas giant planets scales as a power law function of the host star metallicity.

The host star metallicity correlation studies for smaller planets ($R_\mathrm{P}$ $\leq$ 4$\, R_\oplus$) began only after the launch of the $Kepler$ mission because such studies required a large number of smaller planet detections \citep[e.g.,][]{buchhave12,mann13,batalha13,adibekyan13,fressin13,buchhave14,buchhave15,mulders16,CKS4}. These studies showed that the average metallicity of the host stars of small planets ($R_\mathrm{P}$ $\leq$ 4$\, R_\oplus$) is lower than the average metallicity of stars hosting giant planets. Further, the overall scatter in the distribution of metallicity of the host stars was larger for those harboring small planets compared to those hosting giant planets, indicating that the small planets can form around stars with a wide range of metallicities \citep[e.g.,][]{buchhave12,mann13,buchhave14,buchhave15,schlaufman15,wang15,CKS4,mulders18}.

Studies have also shown the presence of a correlation between the metallicity of the host star and the orbital period of the planet around it. For planets orbiting with periods of 10 days or less, the host stars appear to be metal-rich and have higher average metallicity than host stars with planets orbiting farther out \citep[e.g.,][]{beauge13,adibekyan13,mulders16,maldonado17b,wilson17, CKS4}.

In this paper, we investigate the correlations between host star metallicity and planetary radius and orbital period using the latest exoplanet data from $Kepler$, Data Release 25 \citep[hereafter DR25;][]{thompson17}. We supplement the DR25 catalog with improved planetary and stellar radius estimates based on $Gaia$ DR2 from \cite{berger18}. 

An advantage of the large and uniform DR25 $Kepler$ sample is that the observational biases and selection effects are well understood and can be corrected for by calculating the occurrence rate as a function of host star metallicity and planetary properties.  

In Section 2, we describe our sample and the selection criteria used in compiling the sample. In Section 3, 4, and 5, we investigate how the host star metallicity is related to planetary radius and mass.  In Section 6, we discuss the relationship between the orbital period of the planet and host star metallicity. In Section 7, we compare our results obtained using the metallicities from the DR25 stellar catalog \citep{mathur17} with those of \cite{CKS4}, who used more precise metallicities measured from high resolution (R $\sim$ 60,000) spectra of host stars. We show that our results are consistent with those of \cite{CKS4} and summarize them in Section 8.

\section{Sample selection}

To ensure a uniform sample with well understood selection biases we used the latest $Kepler$ Data Release 25 \citep{thompson17} (retrieved on 2018 March 18) from the NASA Exoplanet Archive\footnote{https://exoplanetarchive.ipac.caltech.edu/} \citep{akeson13}. Detailed characterization of the host star properties is important to derive accurate planet properties and to understand the planet population. \cite{huber14} and later \citep{mathur17,mathur18} compiled a catalog of all the stars observed by the $Kepler$ Mission in Quarter 1-16 (Q1-Q16) and Quarter 1-17 (Q1-17; DR25) respectively. These catalogs compiled the host star properties (temperature, surface gravity, and metallicity) from various surveys. For the DR25 Stellar Catalog, \cite{mathur17} used results mainly from LAMOST \citep{lou15}, APOGEE \citep{maj15}, Kepler community follow-up program (CFOP) (spectroscopy), and KIC photometry \citep{brown11}. The temperature, surface gravity, and metallicity were then homogeneously fitted to grids of Dartmouth stellar isochrones to derive stellar properties following \cite{Serenelli13}. \cite{mathur17} then reported the best-fit value conditioned on the isochrone fitting as the stellar parameters for the DR25 catalog. The fit to isochrones is done to derive the stellar radius and age; as a result of this fitting procedure, the effective temperature, surface gravity, and metallicity are changed somewhat from the input values, since they have to be consistent with the values of the isochrones. For our analysis, we replaced the stellar  and planetary radii in DR25 with improved estimates based on $Gaia$~DR2 from \cite{berger18}. The stellar radii estimates from $Gaia$~DR2 are a factor of 4-5 better than previous estimates, which translates into a similar improvement in the planetary radii \citep{berger18}.

In this paper, we focus on planet candidates (identified by the DR25 transit detection run)  around main-sequence stars with spectral type F, G, K, and M with a \teff range of 3200 K (M4) - 7200 K (F0). To ensure that we only pick main sequence dwarfs, we restrict the sample to \lg (in cgs units) values between 4 and 5.  We also restrict the sample to an orbital period of 1 to 365 days. 

On further examining our sample, we found that many $Kepler$ planet candidates have radii $> 20~R_{\oplus}$. The theoretical mass-radius relations derived from models without special inflation mechanisms suggest that the upper limit of planet radii should be close to about 12$\,R_\oplus$ \citep[e.g.,][]{bodenheimer03,fortney07,seager07,chabrier08,mordasini12b,swift12,mordasini15}. Planets with radii up to $\sim17-19$\Re can be explained as inflated Jupiters \citep[e.g.][]{fortney07,Miller09,Demory11,hartman12,Hartman16,LilloBox16,Barros16b,Spake16,Thorngren18}. Any object with a radius above 22-25\Re is unlikely to be a planet.

\begin{figure}
\begin{center}
\epsscale{1.25}
\plotone{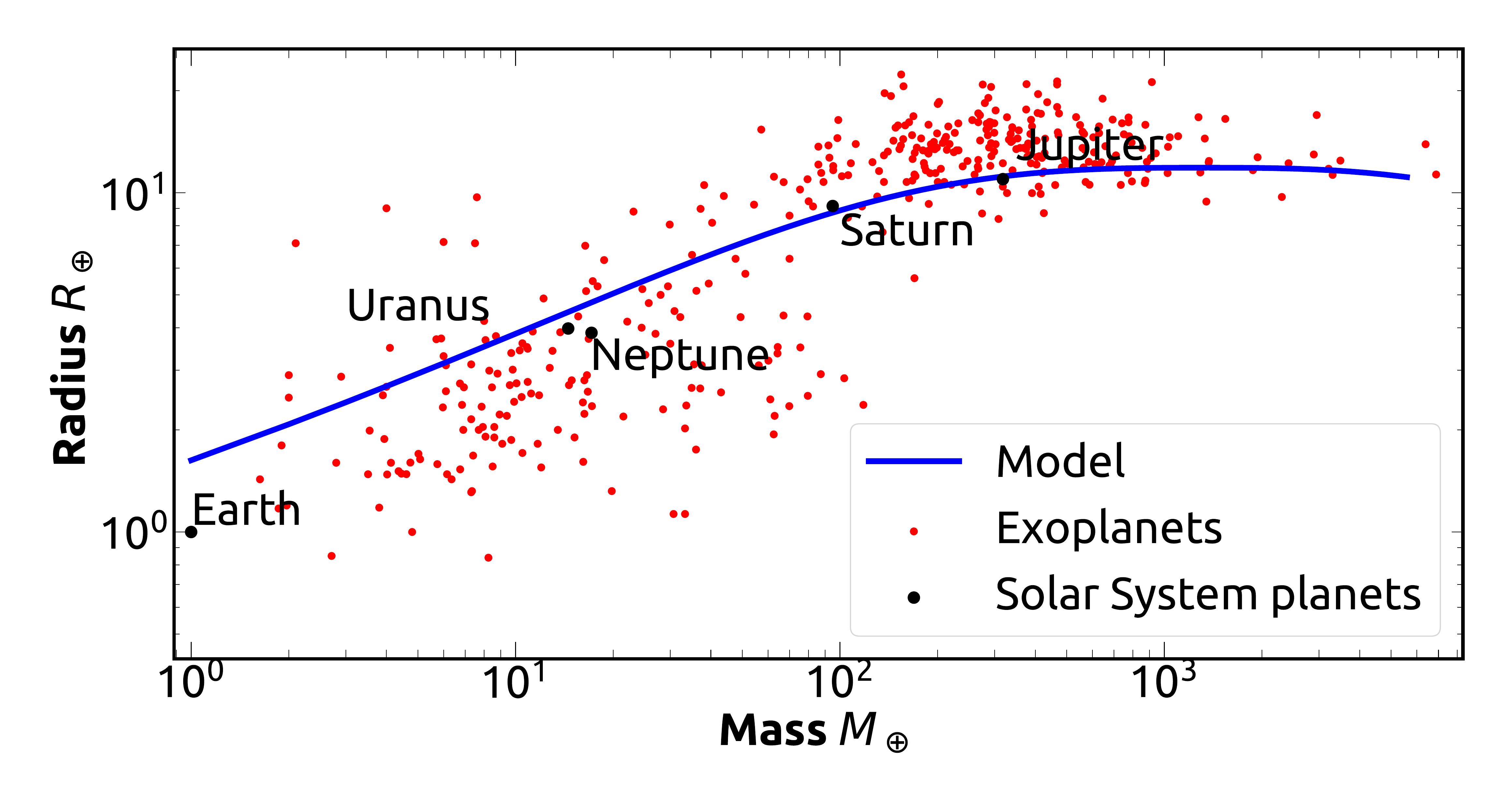}
\caption{The masses and radii for 403 planets (solid red circles). Also plotted are the Solar system planets (black solid circles). We also show the theoretical mass-radius curve (without inflation) from \cite{mordasini12b} in blue. The median uncertainty in the mass measurements of the exoplanets is about 11$\%$ and median uncertainty in radius measurements is about 5$\%$. }

\label{Figure1}

\end{center}
\end{figure}

In Figure \ref{Figure1}, we show the mass-radius relationship for 403 exoplanets (red points) for which both mass and radius are measured independently. These data are obtained from the confirmed planet table at the  NASA Exoplanet Archive. The radii of the planets were measured via transit observations and the masses of the planet were obtained from radial velocity or transit timing variation measurements. In Figure \ref{Figure1}, we have also shown the theoretical fit to the mass-radius relationship (5 Gyr) proposed by \cite{mordasini12b} (blue curve). We clearly see that for Jovian-mass planets the theoretical radius limit is about 12\Re and the measured radii are mostly less than about 20$\, R_\oplus$. Therefore, in our analysis, we restricted ourselves to planets with radii in the range $ 1\,R_\oplus\leq R_\mathrm{P} \leq 20\,R_\oplus$.

\begin{figure}
\begin{center}
\epsscale{1.25}
\plotone{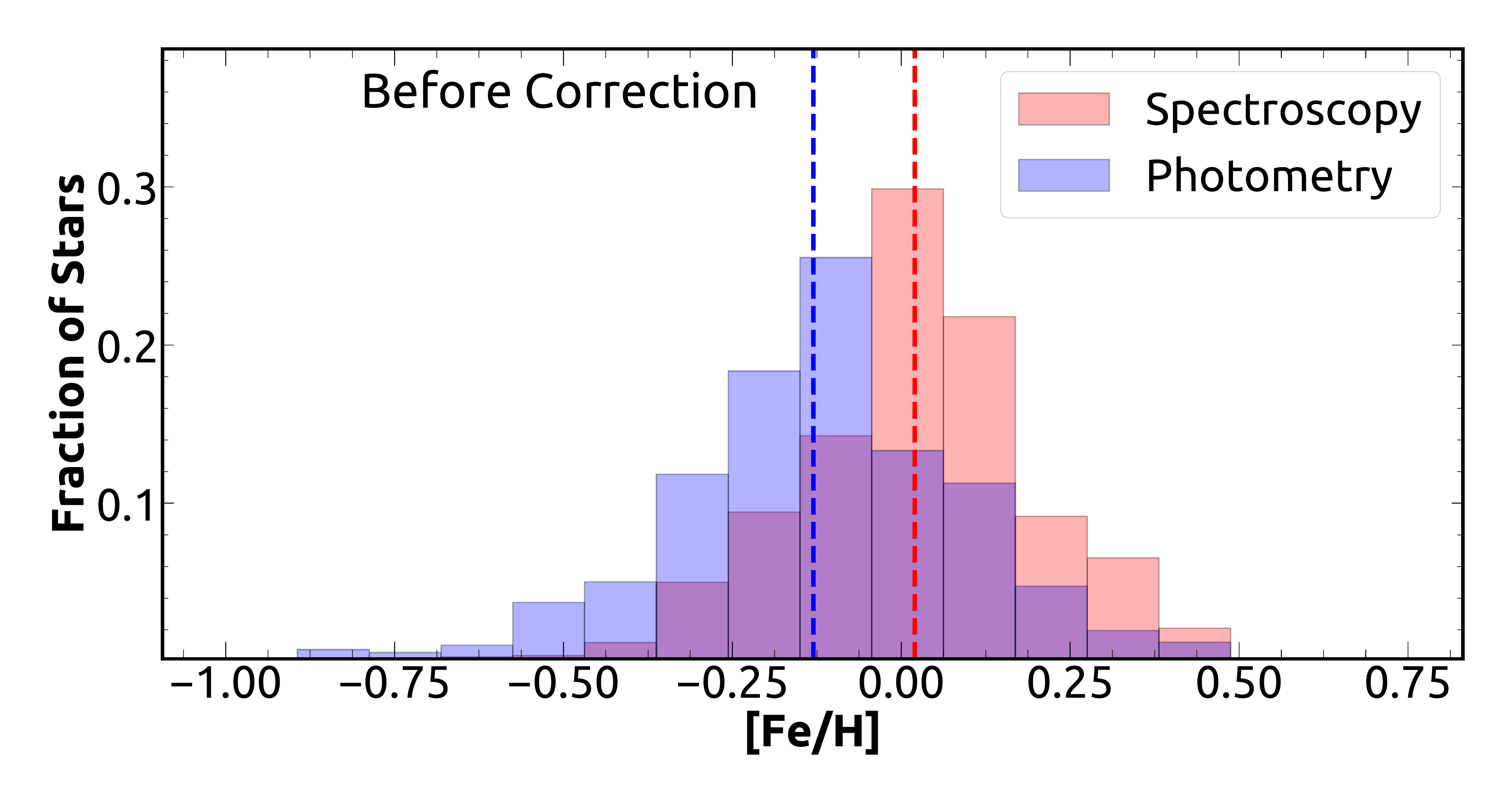}
\plotone{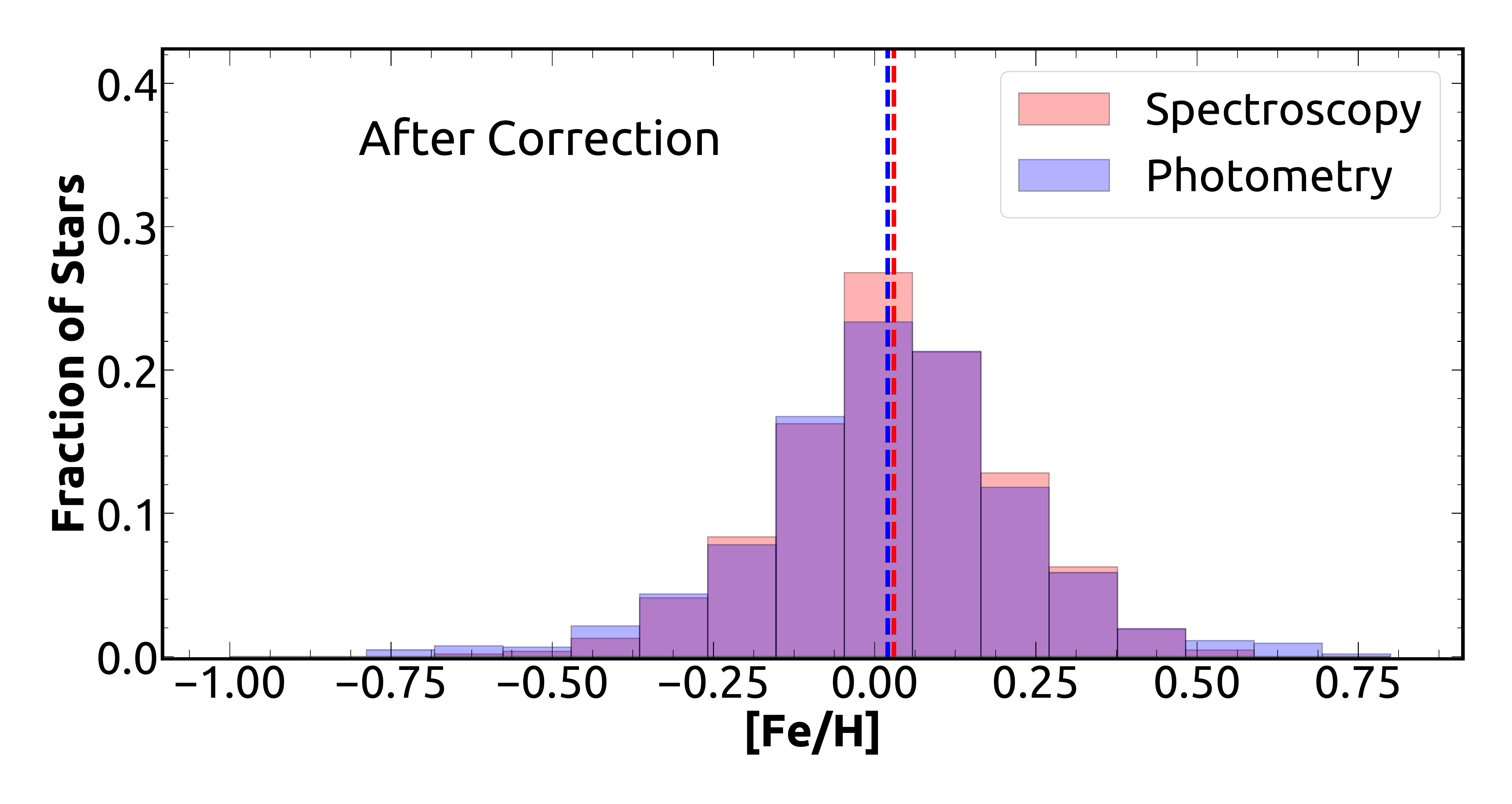}
\caption{Distribution of spectroscopic and photometric host star metallicities from DR25 before (top panel) and after correcting (bottom panel) for the metallicity offset  as described in the text. The dashed blue line represents the mean metallicity of the photometric sub-sample and the red dashed line represents the mean metallicity of the spectroscopic sub-sample.}

\label{Figure2}

\end{center}
\end{figure}

The stellar parameters derived from spectroscopy are relatively more accurate than those derived from photometry, in particular, the metallicity. In the DR25 sample, about 60$\%$ of the host stars of the planet candidates have spectroscopic metallicities. We compared the spectroscopically determined metallicities with those determined photometrically for the planet host stars for our DR25 sample. There appears to be an offset between the two even though the host stars in both the samples have similar spectral type and \lg distribution. This is illustrated in Figure~\ref{Figure2}, where we show the distribution of spectroscopic and photometric metallicities for the planet host stars from the DR25 catalog. As can be seen from Figure~\ref{Figure2}, the mean of the spectroscopic and photometric metallicity distributions are offset by $\sim$~0.15 dex.

 In order to better characterize this offset and to correct for it, we compared the DR25 host star metallicities with metallicities from California Kepler Survey (CKS) \citep{CKS1,CKS2,CKS4} and LAMOST Data Release 4~(DR4)\footnote{http://dr4.lamost.org/} for sources that are common between DR25 and these two catalogs. We find that the offset between the DR25 spectroscopic metallicities and CKS metallicities~(for 640  common sources) is~$\sim$~0.01~dex. The  offset between the DR25 spectroscopic metallicities and LAMOST~DR4 metallicities (for 482 common sources) is~$\sim$0.03~dex. On the other hand, the DR25 photometric metallicities show  offsets of 0.15~dex compared to CKS metallicities (168 common sources) and 0.14~dex compared to LAMOST~DR4 metallicities (308 common sources). In order to make the DR25 spectroscopic and photometric metallicities consistent with each other and with those in CKS and LAMOST, we applied a correction of 0.15~dex to photometric metallicities and 0.01~dex to spectroscopic metallicities listed in the DR25 catalog. The metallicity distribution of the DR25 host star sample after applying the correction described above is shown in Figure~\ref{Figure2} (bottom panel). Both the spectroscopic and photometric metallicity distributions look very similar after applying the offset correction.  

 We further carried out a similar comparison of metallicities for the full DR25 stellar sample with metallicities listed in the LAMOST~DR4 for sources common between the two. The  offsets found between DR25 spectroscopic metallicities and LAMOST~DR4 metallicities (for 4435 common sources) and DR25 photometric metallicities and LAMOST~DR4 metallicities (for 27551 common sources) are very similar to those found for planet host star metallicities. Therefore, we applied the same corrections to the spectroscopic and photometric metallicities for larger DR25 stellar sample as well. We use these corrected metallicities in our analysis.  In this analysis we restrict ourselves to stellar metallicities between -0.8 and 0.5.

After applying all these filters, our final  sample contains a total of 2864 $Kepler$ planet candidates around 2142 main sequence F, G, K, and M type stars. This is the largest sample to date for which the dependence of observed planet properties (radius and orbital period) on host star metallicity has been studied.

\section{Planet radius and host star metallicity}
\begin{figure*}
\begin{center}
\epsscale{1.25}
\plotone{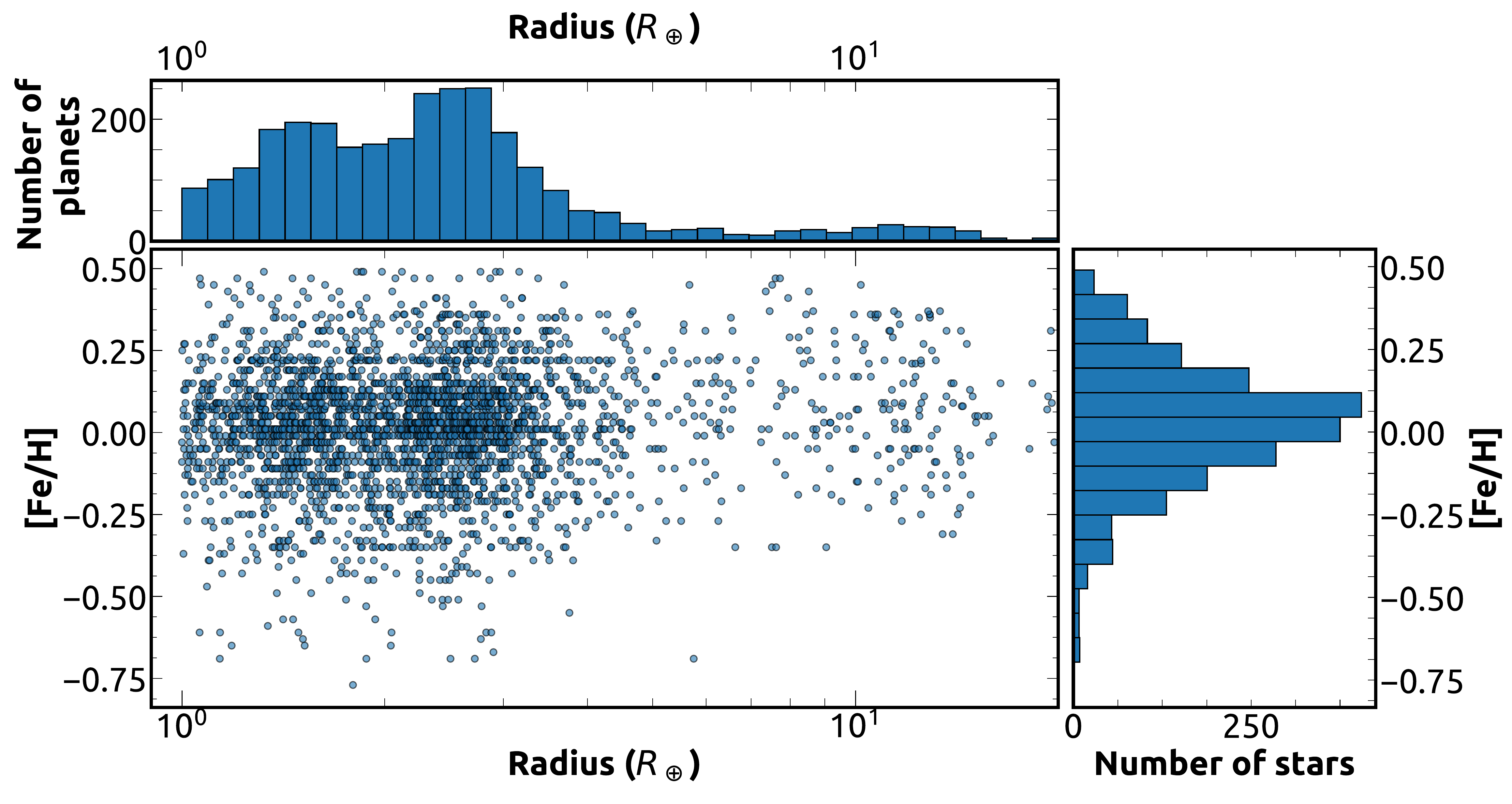}
\caption{A scatter plot between the planetary radius and the host star metallicity for the $Kepler$ DR25 sample. Also shown are the planet radius and host star metallicity distribution on the top and to the right. The host star metallicity histogram list each host star only once so as to avoid over-counting for multi-planetary systems (this convention is followed in all subsequent plots).}
\label{Figure3}

\end{center}
\end{figure*}

\begin{figure}
\begin{center}
\epsscale{1.25}
\plotone{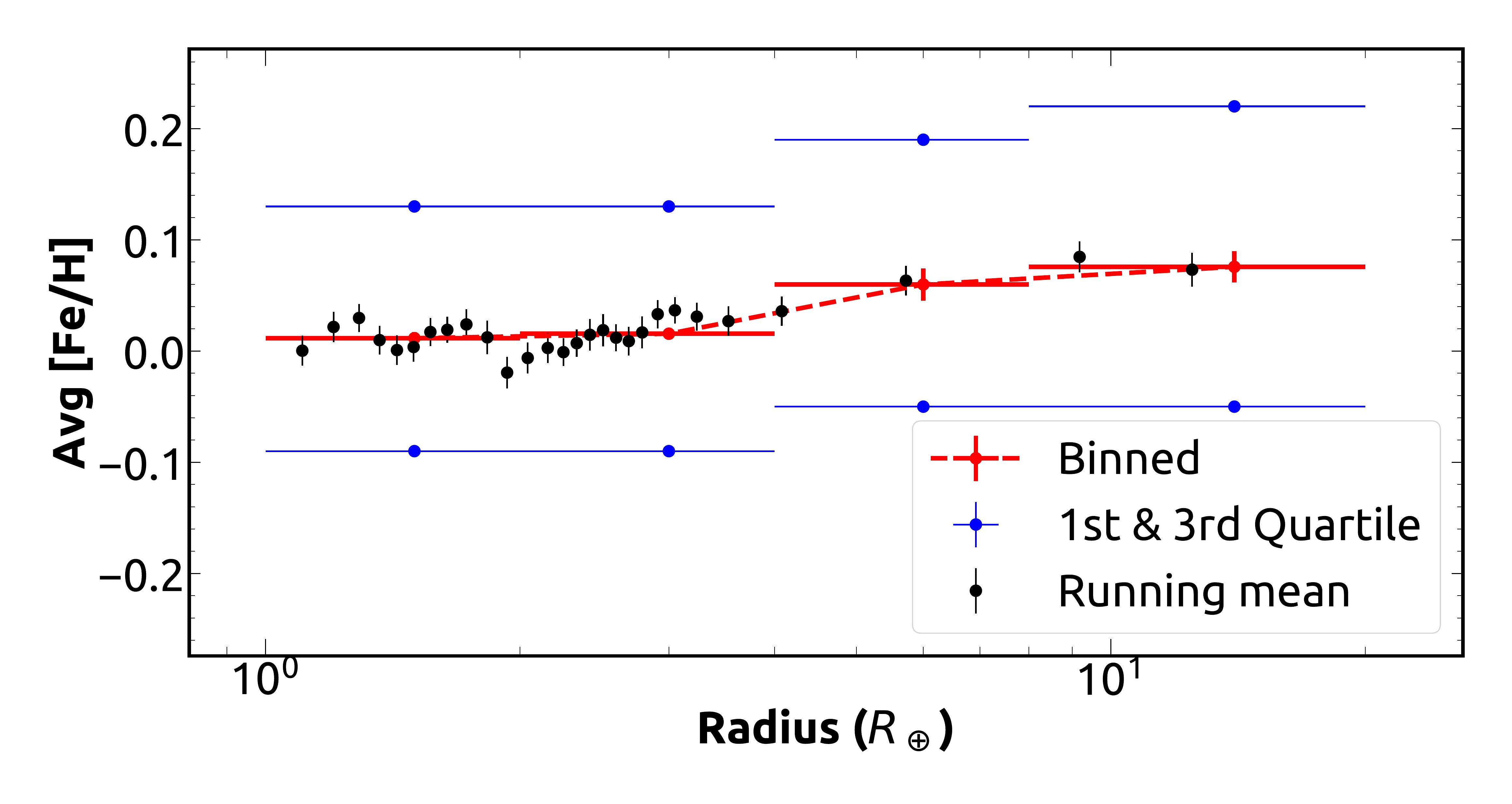}
\caption{Average host star metallicity as a function of planet radius for the planetary candidate sample (2864 planet candidates in total). The red solid circles are the mean values of [Fe/H] and radius in each bin. The width of the radius bins is shown as red horizontal lines. The red vertical bars represent the standard deviation of the mean in [Fe/H] and blue lines represent upper and lower quartiles for each bin. We also show a running mean of the host star metallicity as a function of the planetary radius computed for a box size of 200 with a step size of 100 as black solid circles with the vertical (black) lines representing the standard deviation of the mean in [Fe/H].}
\label{Figure4}

\end{center}
\end{figure}
\subsection{Planet classification based on radius}

Exoplanets come in various radii and masses. Based on their radii and masses several classifications have been suggested in the literature \citep[e.g.,][]{borucki12,adibekyan13,fressin13,buchhave14,CKS4}. \cite{CKS3} have shown that the occurrence rate of planets as a function of radius has a gap at $1.7\,R_\oplus$; however, the improved radius estimates based on Gaia~DR2 from \cite{berger18}, show that the gap is at $\sim 2\,R_\oplus$. Keeping this in mind, we classify planets in our sample as super-Earths $(1\,R_\oplus\leq R_\mathrm{P} \leq 2\,R_\oplus),$ Neptunes $(2\,R_\oplus < R_\mathrm{P}\leq 4\,R_\oplus)$, sub-Saturns $(4\,R_\oplus <R_\mathrm{P} \leq 8\,R_\oplus)$  and Jupiters $(8\,R_\oplus <R_\mathrm{P} \leq 20\,R_\oplus)$. This binning scheme is quite similar to that of \cite{CKS4} with two small differences: the cutoff point between super-Earths and Neptunes is set at 2\Re and the upper limit of planetary radius is set to  $20\,R_\oplus$. 

\subsection{Planet radius and host star metallicity from the DR25 stellar catalog}
Figure \ref{Figure3} shows the distribution of the \plr and \hsm for the DR25 sample. The planetary radii are from \cite{berger18} and the host star  metallicities (corrected) are from the DR25 stellar catalog \citep{mathur17,mathur18} as described earlier. In Figure \ref{Figure4}, we show host star average metallicity as a function of planet radius. We binned the data in the radius bins described in Section 3.1. Figure \ref{Figure4} shows that although there is large scatter in [Fe/H] in each radius bin, the average host star metallicity rises as the planetary radius increases, indicating that larger planets are preferentially found around host stars with higher metallicity. 

\begin{figure}
\begin{center}
\epsscale{1.25}
\plotone{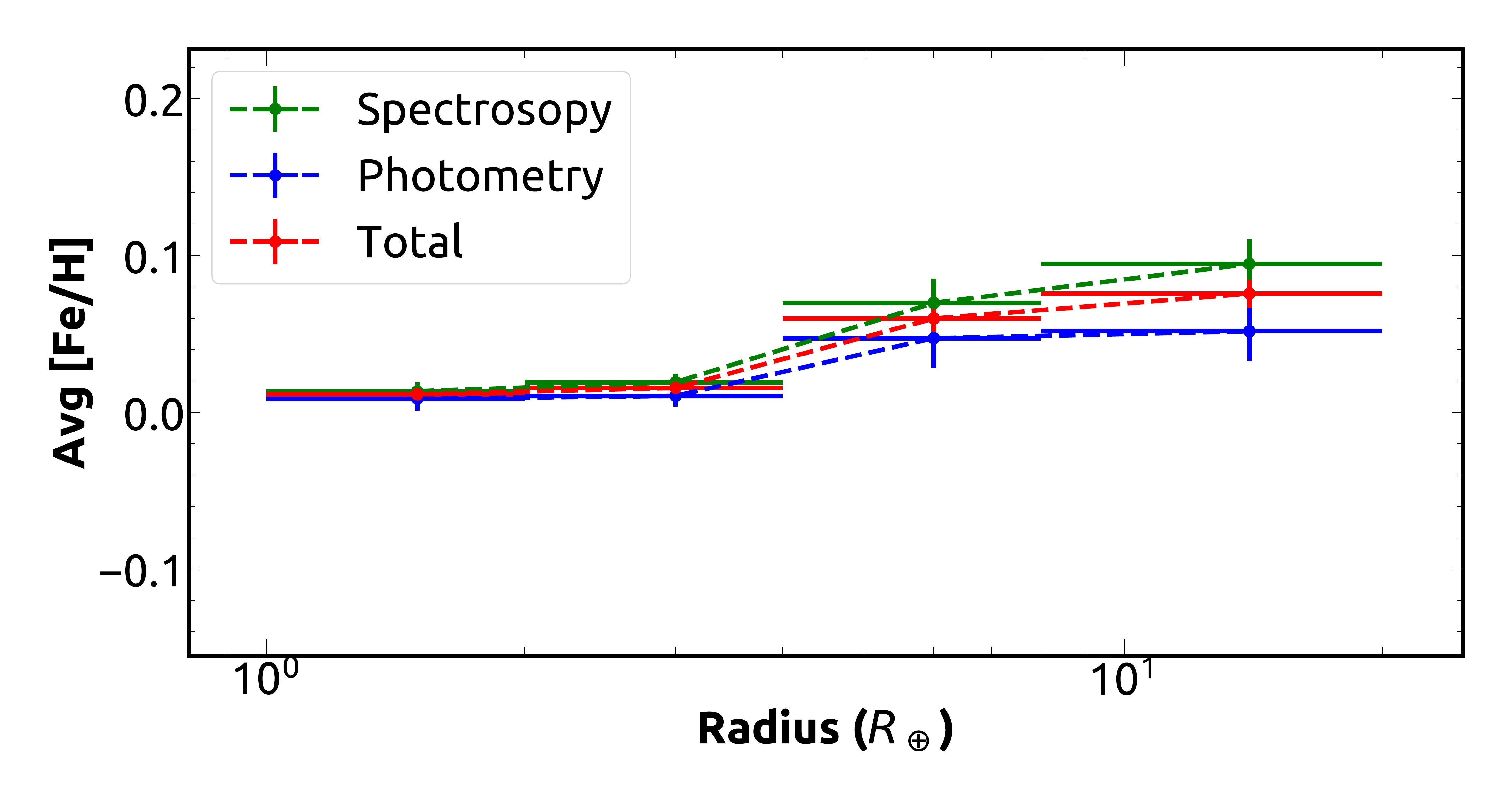}
\caption{The average host star metallicity planet radius relation  obtained after 100,000 random realizations of host star metallcities. The median value of the average metallicity of each bin for 100,000 realizations is shown as a function of planet radius. The standard deviation in the average metallicity of each bin is shown as the vertical error bar. The blue curve represents the average host star metallicities measured via photometry, the green curve represents host star metallicities measured via spectroscopy, the red curve represents the total sample. The binning scheme is the same as in Figure~\ref{Figure4}.   }
\label{Figure5}

\end{center}
\end{figure}

Although we have applied the correction to the DR25 metallicities to make the spectroscopic and photometric metallicities consistent with each other, the photometric metallicities in the DR25 catalog have relatively large uncertainties (mean uncertainty~$\sim$~0.28 dex) compared to spectroscopically determined metallicities (mean uncertainty~$\sim$~0.14 dex). We next checked to see if these uncertainties affect the robustness of the statistical results that we find in Figure~\ref{Figure4}. To do so we created 100,000 realizations of the metallicity distributions of the DR25 planetary sample with the metallicity (corrected) of each candidate host star chosen randomly such that the metallicity lies within the uncertainty of the measurement. Then we binned the data in radius bins exactly the same way as was done in Figure~\ref{Figure4}. The average metallicity in each bin was also calculated separately for host stars with spectroscopic and photometric metallicities. We then plot the median value of the average metallicity of each bin for the 100,000 realizations as a function of planet radius. This is displayed in Figure~\ref{Figure5}.   As can be seen from the figure, both the spectroscopic and photometric samples show a similar trend with planet radius. Also, for the total simulated sample, the average metallicity exhibits almost the same behavior as that in Figure~\ref{Figure4}. This indicates that even though the uncertainties in individual metallicities could be relatively high, particularly for photometric metallicities, the overall statistical correlation that we find between average host star metallicity and the planetary radius is robust.

\subsection{ Planet radius and host star metallicity relationship from the California Kepler Survey }

To make use of the more precisely determined host star properties, we repeated the analysis using the data from the California Kepler Survey (CKS). The stellar properties for CKS have been determined using high-resolution spectroscopy \citep{CKS1, CKS2}.

\begin{figure*}
\begin{center}
\epsscale{1.25}
\plotone{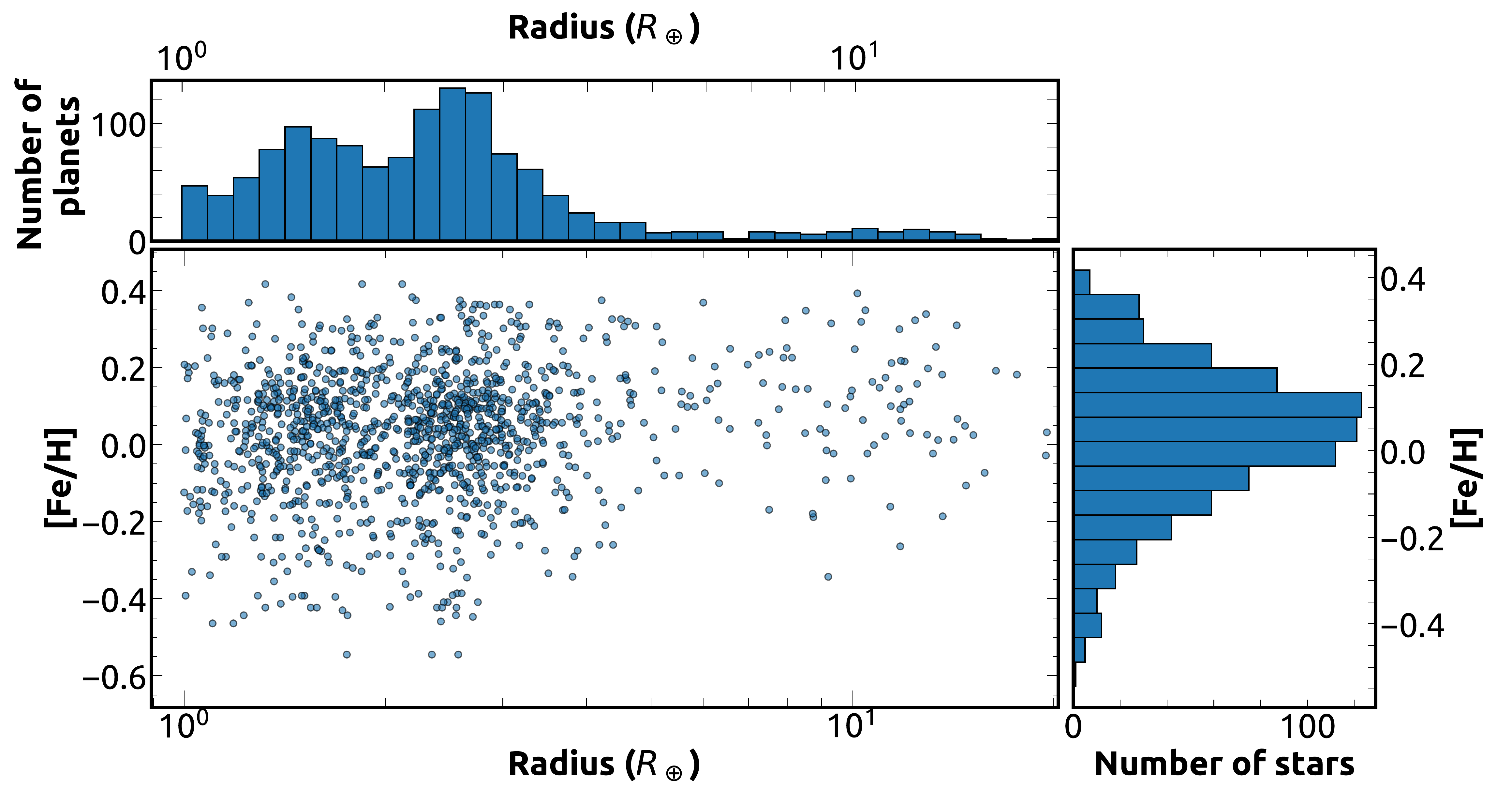}
\caption{A scatter plot between planetary radius and \hsm for the CKS sample. See Figure \ref{Figure3} for a description of the plot. }
\label{Figure6}

\end{center}
\end{figure*}

We examined the correlation between host star metallicities from CKS and planet radii estimates from \cite{berger18} (in order to maintain consistency). Using $Kepler$ DR25 we updated the disposition status of all the planets in the CKS sample and only selected planets that were classified as candidates based on DR25. We then applied the same filters to the CKS sample as those used for our DR25 sample. The \teff range is from 3200 K - 7200 K,  \lg between 4 and 5, and planetary radius between $ 1\,R_\oplus$ and $20\,R_\oplus$ (see Section 2). After applying these filters, there were in total 1317 $Kepler$ candidates with CKS stellar parameters. \cite{CKS4} had a magnitude-limited sample. We do not put any constraints on the magnitude of the host star. We take a much wider \teff range and do not put any constraints on the impact factor of the planet. We also do not exclude planets with host stars that have a nearby stellar companion. Figure \ref{Figure6} shows the distribution of \plr and \hsm for the CKS sample.

\begin{figure}
\begin{center}
\epsscale{1.25}
\plotone{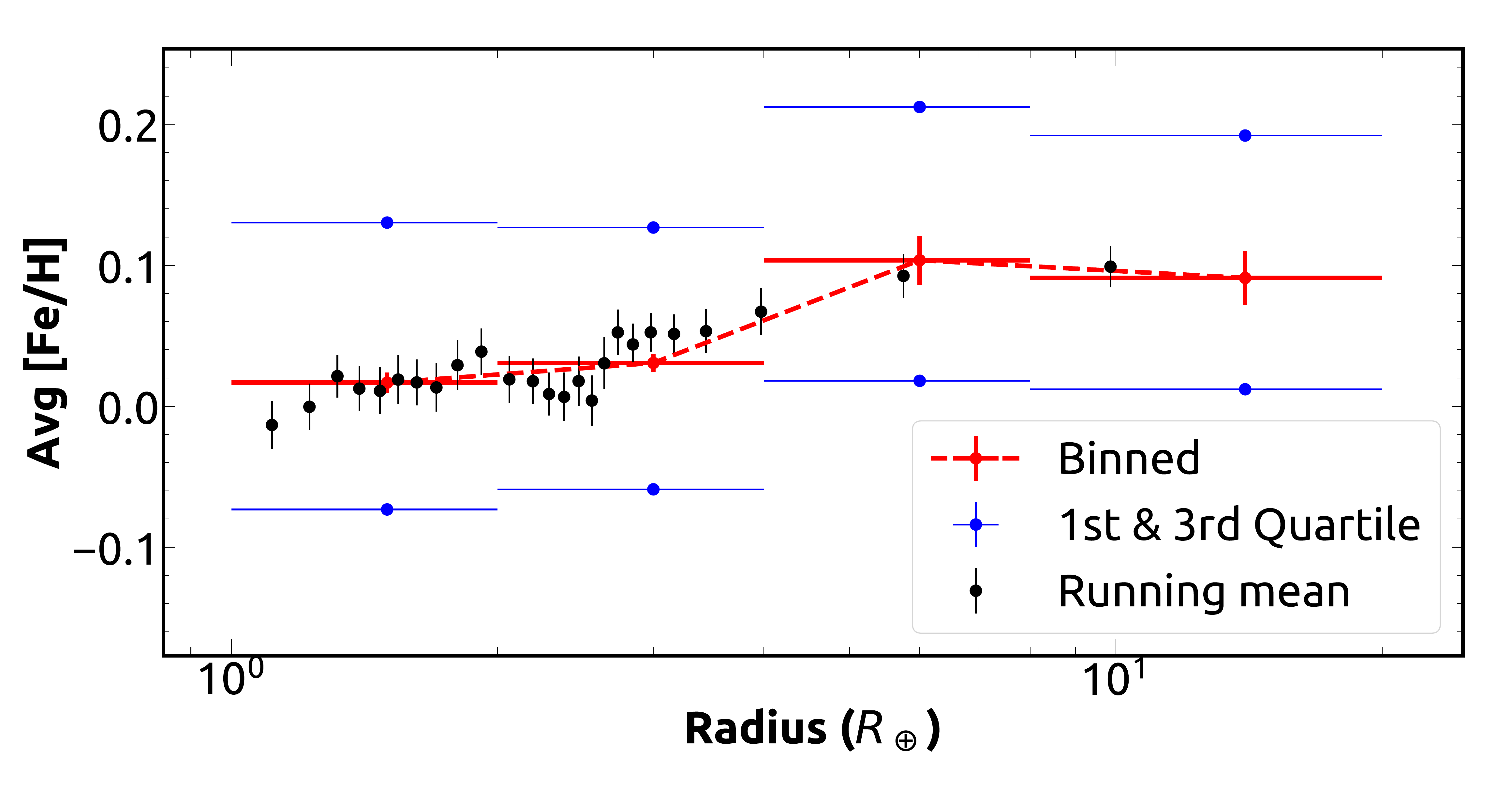}
\caption{Average host star metallicity as a function of planet radius for the 1317 planets selected in the same manner as the DR25 sample, using  stellar metallicities from CKS. The symbols and lines have the same meaning as in Figure \ref{Figure4}. }
\label{Figure7}

\end{center}
\end{figure}

In Figure \ref{Figure7}, we show host star metallicity as a function of planet radius for the CKS sample. Here again, we binned the data into various radius bins as described in Section 3.1. We follow the same convention for plotting as in Figure \ref{Figure4}. For the running mean, we used a box size of 100 with a step size of 50. The results from Figure \ref{Figure7} are consistent with Figure \ref{Figure4} and also with the results presented in \cite{CKS4}.

\section{Planet mass and host star metallicity}

Studies prior to $Kepler$ have shown that giant planets are preferentially found around metal-rich host stars \citep[e.g.,][]{gonzalez97,santos01,fischer05,johnson10}. Recently, \cite{santos17} found that, on average, the host star metallicity of more massive giant planets $(M_\mathrm{P} > 4\,M_\mathrm{J})$ is lower than the host star metallicity for less massive giant planets with $(M_\mathrm{P} \leq 4\,M_\mathrm{J})$ (also see \cite{schlaufman18}). 
 
\begin{figure}
\begin{center}
\epsscale{1.25}
\plotone{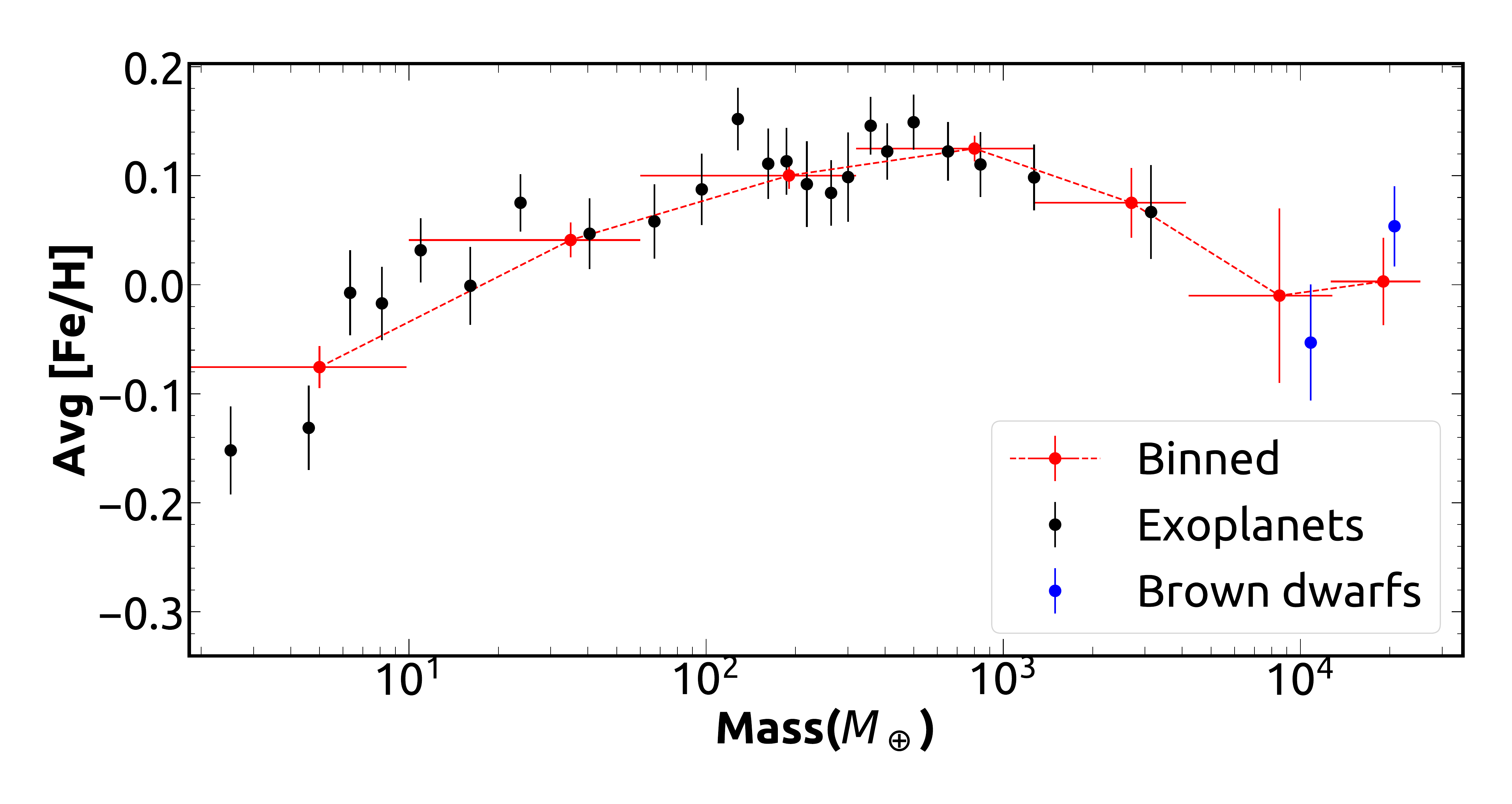}
\caption{Average host star metallicity as a function of planet/brown dwarf mass. The average host star metallicity for various (companion) mass bins is shown in red with the horizontal red line representing the bin width and the vertical red line the standard deviation of the mean. We also show a running mean of the host star metallicity as a function of the planetary mass computed for a box size of 60 with a step size of 30 as black solid circles with the vertical (black) lines representing the standard deviation of the mean in [Fe/H]. We also show the average metallicities for host stars with brown dwarfs companions (blue) with vertical (blue) lines representing the standard deviation of the mean in [Fe/H] of the host stars of brown dwarfs. } 
\label{Figure8} 

\end{center}
\end{figure} 
 
To study the relationship between \hsm and planetary mass we retrieved data from the confirmed planets table from the NASA Exoplanet Archive (retrieved on 11 May 2018). The catalog lists upper limits for some of the planetary masses, so we searched the literature for better mass estimates and replaced them. When only the upper limits were available, we did not include the planet in the sample. Planets with radial velocity (RV) measurements and inclination angle measurements allow us to derive the true mass \citep[e.g.,][]{batalha11,marcy14,gettel16}. If the inclination angle was not available $M\mathrm{sin}i$ the minimum mass, was used as a proxy of planet mass. We also included planets with masses determined from transit timing variation (TTV) studies \citep[e.g.,][]{weiss13,hadden14} to increase the sample size.

The stellar parameters $T_\mathrm{eff}$, log$\,g$, and metallicity were taken from the SWEET-Cat catalog \citep{santos13}. The SWEET-Cat catalog derives the host star properties from high-resolution spectra. Care has been taken while maintaining and updating this catalog to make sure that the determination of stellar properties is as uniform as possible. The stellar parameters listed in the NASA Exoplanet Archive are compiled from various different sources. We, however, wanted to use a uniform set of metallicity values for our analysis; hence we opted for the SWEET-Cat catalog for the stellar parameters. We cross-matched the positions of the host stars from the confirmed planet table from the NASA Exoplanet Archive to those listed in the SWEET-Cat catalog with a search radius of 30 arcseconds. We then applied the same filters as those used for the $Kepler$ DR25 sample: \hs \teff range of 3200 K - 7200 K and \hs \lg between 4 and 5 and the orbital period of the planet between 1 and 365 days. After applying these filters, we had a total of 705 confirmed planets with masses listed in the catalog. Out of these, 175 planets are in the $Kepler$ field and 109 have TTV measurements.

In Figure \ref{Figure8}, we show the average host star metallicity as a function of planetary mass. 
We also include binary brown dwarfs along with the metallicities of their host stars in Figure \ref{Figure8}. The brown dwarf data is from \cite{ma14}. We applied the same filters that we applied for the $Kepler$ sample.  The average \hsm for planets/brown dwarfs in various mass bins (1$~M_\oplus$ - 10$~M_\oplus$, 10$~M_\oplus$ - 50$~M_\oplus$, 50$~M_\oplus$ - 1$\,M_\mathrm{J}$, 1$\,M_\mathrm{J}$ - 4$\,M_\mathrm{J}$, 4$\,M_\mathrm{J}$ - 13$\,M_\mathrm{J}$, 13$\,M_\mathrm{J}$ - 35$\,M_\mathrm{J}$, and 35$\,M_\mathrm{J}$- 80$\,M_\mathrm{J}$) is shown as red solid circles. The brown dwarfs are shown separately as blue circles with the blue vertical bars being the standard deviation in the mean of [Fe/H] of their host star. 

As in the case for the radius-metallicity relation, the average host star metallicity increases as the mass of the planet increases until about 4\Mj ($\sim 1200\, M_\oplus$). For planet masses $> 4 \, M_\mathrm{J}$, the trend appears to reverse: the \hsm begins to drop as the mass of the planet increases. Figure \ref{Figure8} clearly shows that the drop in the metallicity of the host star as the companion mass increases extends into the brown dwarf regime as well. Similar results have also been reported by \cite{santos17} and \cite{schlaufman18} for a sample of binaries with giant planets, brown dwarfs and low mass stars as secondary companions. These results suggest a similar mechanism for the formation of super-Jupiter $(> 4\, M_\mathrm{J})$, brown dwarf and low mass star binary systems.

From simulations, it has been shown that for Jupiters ($1-4 \, M_\mathrm{J}$), [Fe/H] is not the only parameter that determines the final planet mass. [Fe/H] determines whether or not a giant planet can form, but not the mass \citep{mordasini12}. The mass is rather correlated with the disk gas mass, not [Fe/H]. For very massive giant planets ($\geqslant$ 10 $M_\mathrm{J}$), this is no more true: in order to become a very massive giant planet, the critical core mass must form very fast, before the gas in the disk is dissipated by accretion onto the star or photo-evaporation \citep{mordasini12}. This is only possible at high [Fe/H]. In fact, \cite{mordasini12} based on their theoretical planet population models found that the most massive giant planets ($\geqslant$ 10 $M_\mathrm{J}$) only form at high [Fe/H]. This prediction of the core-accretion model is quite contrary to what we find. This may indicate another formation mechanism, e.g.,  gravitational instability, for the formation of super-Jupiters ($M_\mathrm{P} >$ 4 $\, M_\mathrm{J}$).

\section{Occurrence Rate}

The analysis described in the previous sections does not take the completeness of the survey or the detector efficiency into account. The true trend might not be the one which simple binning or a running average shows. In order to derive the correlation between host star metallicity and the planet size that is free of observational biases and other selection effects, we use the latest $Kepler$ data release DR25 catalog to calculate the occurrence rate of exoplanets as a function of radius and metallicity. We calculate the occurrence rates using the methods described extensively in the literature \citep[e.g.,][]{youdin11,howard12,dressing13,fressin13,morton14,burke15,mulders15drop,mulders16}. 

\subsection{Number of stars with detectable planets}

Whether we can detect a planet around a given star depends on various factors: the geometric probability of the planet transiting the host star along our line of sight; the signal to noise that the transit would produce, which, in turn, is a factor of the transit depth, $\sigma_{CDPP}$, and the number of transits. In the following subsections, we will describe how these quantities are calculated.
 
\subsubsection{Transit probability}
 The transit probability is defined as the geometric probability that we will be able to observe a given transit of a planet around a star of radius $R_*$ at an orbital distance or semi-major axis $a$.

Using simple geometry one can derive the transit probability $\eta_{tr}$ as 
\begin{equation*}
\eta_{tr}=\frac{R_*+R_\mathrm{P}}{a}\frac{1}{1-e^2}\approx \frac{R_*}{a(1-e^2)}
\end{equation*}

Here $R_*$ is the radius of the host star, $R_\mathrm{P}$ is the radius of the planet, $e$ is the eccentricity, and $a$ is the semi-major axis of the planet. Here we assume negligible eccentricity (i.e. circular orbits) and further approximative $\eta_{tr}$ as

\begin{equation}
\eta_{tr}= \frac{R_*}{a(1-e^2)} \approx \frac{R_*}{a}
\label{1}
\end{equation}

Since $a$, the semi-major axis is not an observable we convert $a$ into the orbital period of the planet around the host star using Kepler's $3^{rd}$ law and stellar mass $M_*$ taken from the stellar catalog. 

\begin{equation}
\eta_{tr}\approx\frac{R_{*}}{a}=R_{*}\big(\frac{4\pi^{2}}{GM_{*}P^{2}}\big)^{1/3} 
\label{2}
\end{equation}

\subsubsection{Multiple Event Statistic}
The Multiple Event Statistic is a measure of how reliable the transit detection is. It is a measure of how noisy the signal would be, given that our planet of radius $R_\mathrm{P}$ orbits a star with radius $R_*$ with a period of $P$ days, provided the star has a noise level $\sigma_{CDPP(\tau)}$ (combined differential photometric precision interpolated to transit duration). $\sigma_{CDPP}$ can be described as the effective white noise that a transit signal would see given the signal has a duration of $\tau$. A $\sigma_{CDPP}$ of 30 ppm for a 1-hour transit indicates that a 1-hour transit of depth 30 ppm would on average have an SNR of 1 \citep{christiansen12}. The SNR depends on the transit depth, transit duration, $\sigma_{CDPP(\tau)}$ and the number of transits.

The transit depth $\delta$ is the fraction of flux blocked by the planet $\Delta F$ to the total flux $F$ emitted by the star and is given by 

\begin{equation}
\frac{\Delta F}{F}=\delta=\frac{\mathrm{{Area\,of\,planet}}}{\mathrm{Area\,of\,star}}=\frac{R_{P}^{2}}{R_{*}^{2}}
\label{3}
\end{equation}

Transit duration, the time duration in which any part of the planet obscures the disk of the star, is given by 
\begin{equation*}
\tau_{dur}=6\bigg(\frac{P}{\mathrm{1~ day}}\bigg)\;\bigg(\frac{R_{*}}{a}\bigg) \sqrt{{1-e^2}}\,hr 
\end{equation*}

\begin{equation}
\approx 6\bigg(\frac{P}{\mathrm{1~ day}}\bigg)\;\bigg(\frac{R_{*}}{a}\bigg)\,hr 
\label{4}
\end{equation}
 Since $\sigma_{CDPP}$ is only available as an array of time values for {[}1.5, 2.0, 2.5, 3.0, 3.5, 4.5, 5.0, 6.0, 7.5, 9.0, 10.5, 12.0, 12.5, 15.0{]} hrs we use equation (\ref{4}) \citep{burke15} to calculate the transit duration and interpolate the value of $\sigma_{CDPP}$ to the transit duration to produce the effective white noise on the detector for the transit duration. 
To calculate the overall MES for the transit we also need the total number of transits observed by Kepler during the total quarters observed. But not all the stars were observed for the whole Kepler mission duration. Due to the movement of the of the spacecraft,sometimes some of the stars were not observed as the signal from the stars did not fall on the CCDs. The duty cycle of the star is the ratio of the amount of time the star was observed by the total time period of the Kepler mission. We restrict the sample to stars and KOIs with duty cycle being greater than 60\% and with the star’s data span on the CCD of at least 2 years. Due to these restrictions we calculated the occurrence rate for 2775 Kepler candidates around 2080 main sequence stars.

The total number of transits $N_{tr}$ observed by $Kepler$ during the total quarters observed is given by 

\begin{equation}
N_{tr}=\frac{\mathrm{total\,quarters\,observed*duty\,cycle}}{\mathrm{orbital\,period}}
\label{5}
\end{equation}

We then calculate  MES (Multiple Event Statistic)  by averaging the transit signal
strength over multiple transit events. 

\begin{equation}
MES=\frac{\delta}{\sigma_{CDPP(\tau)}} \sqrt{N_{tr}}
\label{6}
\end{equation}

\subsection{Calculating the occurrence rate}

In order to investigate how the the planetary radius and the orbital period of the planets depend of the host star metallicity, we calculate the planet occurrence rate for three host star metallicity bins: sub-solar ($-0.8 \leq \mathrm{[Fe/H]} <-0.2$), solar ($-0.2 \leq \mathrm{[Fe/H]} \leq 0.2$) and super-solar ($\mathrm{[Fe/H]}>0.2$) (following \cite{beauge13}).

The occurrence rate for planets within a radius bin $R_0$ and period bin $P_0$ is given as \citep{Hsu18}

\begin{equation}
\large F_{{R_0}{P_0}}=\Sigma_{i\,{R_0}{P_0}} C_{i}/N_{targ}
\label{7}
\end{equation}

where $N_{targ}$ is the total number of stars in that metallicity bin and $C_{i}$ is the estimate of the number of planets with radius  $R_0$ and period $P_0$. To account for incompleteness due to the geometric probability of transit and detection efficiency $C_{i}$ is defined as 

\begin{equation}
C_{i} = \frac{1}{\eta_{tr} \eta_{det}}
\end{equation}

$\eta_{det}$ accounts for the fraction of stars around which a planet with  a radius  $R_0$ and period  $P_0$  can be detected.  $\eta_{det}$ times $N_{targ}$ is the effective number of stars around which we can detect such a planet.

\begin{equation}
\eta_{det} =\Sigma_{j=1}^{N_{targ}} \eta_{{R_0},{P_0 }, j} /N_{targ}
\end{equation}

$\eta_{{R_0},{P_0 }, j}$ is the probability that  we can detect a planet with radius $R_0$ and period $P_0$ around the $\mathrm{j^{th}}$  target star. 
Following \cite{burke15} the detection probability from the Kepler pipeline is a function of Pipeline Completeness  and Window Function.

The modeled pipeline detection probability (from the recovery of injected transit signals) is given by

\begin{equation}
P_{gamma} (x|a,b,c)= \frac{c}{b^a \Gamma (a)} \int_{0}^{x} t^{a-1} e^{-t/b}dt
\end{equation}

where $\Gamma$ is the gamma function, a =30.87, b= 0.271,  and c=0.940 (for pipeline version 9.3 (DR25)),  x is the expected MES. \citep{christiansen17}

The Window Function $P_{win}$ accounts for the probability that a requisite number of transits required for detection occurs \citep{burke15}. 

\begin{equation*}
P_{win}=1-(1-f_{duty})^{M}-M f_{duty} (1-f_{duty})^{M-1}
\end{equation*}

\begin{equation}
-\frac{M(M-1)}{2} f_{duty}^{2}   (1-f_{duty})^{M-2}
\end{equation}
where M = $T_{obs}/P_{orb}$ with $T_{obs}$ being the data-span of the star on the detector and $P_{orb}$ being the orbital period of the planets.   $f_{duty}$ is the duty cycle of the star. 

The detection probability around the $\mathrm{j^{th}}$  target star is then given as
\begin{equation}
 \eta_{{R_0},{P_0 }, j}=P_{win} * P_{gamma}
\end{equation}

Equation (\ref{7}) takes into account all the observational biases and gives us a true measure of the occurrence rate. To get the occurrence rate as a function of radius and host star metallicity we simply sum over the period.

\begin{equation*}
F\,\bigg|_{R_{p}=R_{0}} = \Sigma_{i}  {F}_{{R_0}{P_i}}
\end{equation*}

To get the occurrence rate as a function of period and host star metallicity we simply sum over the radius.

\begin{equation*}
F\,\bigg|_{P=P_{0}} = \Sigma_{i} F_{{R_i}{P_0}}
\end{equation*}

\begin{figure}
\begin{center}
\epsscale{1.25}

\plotone{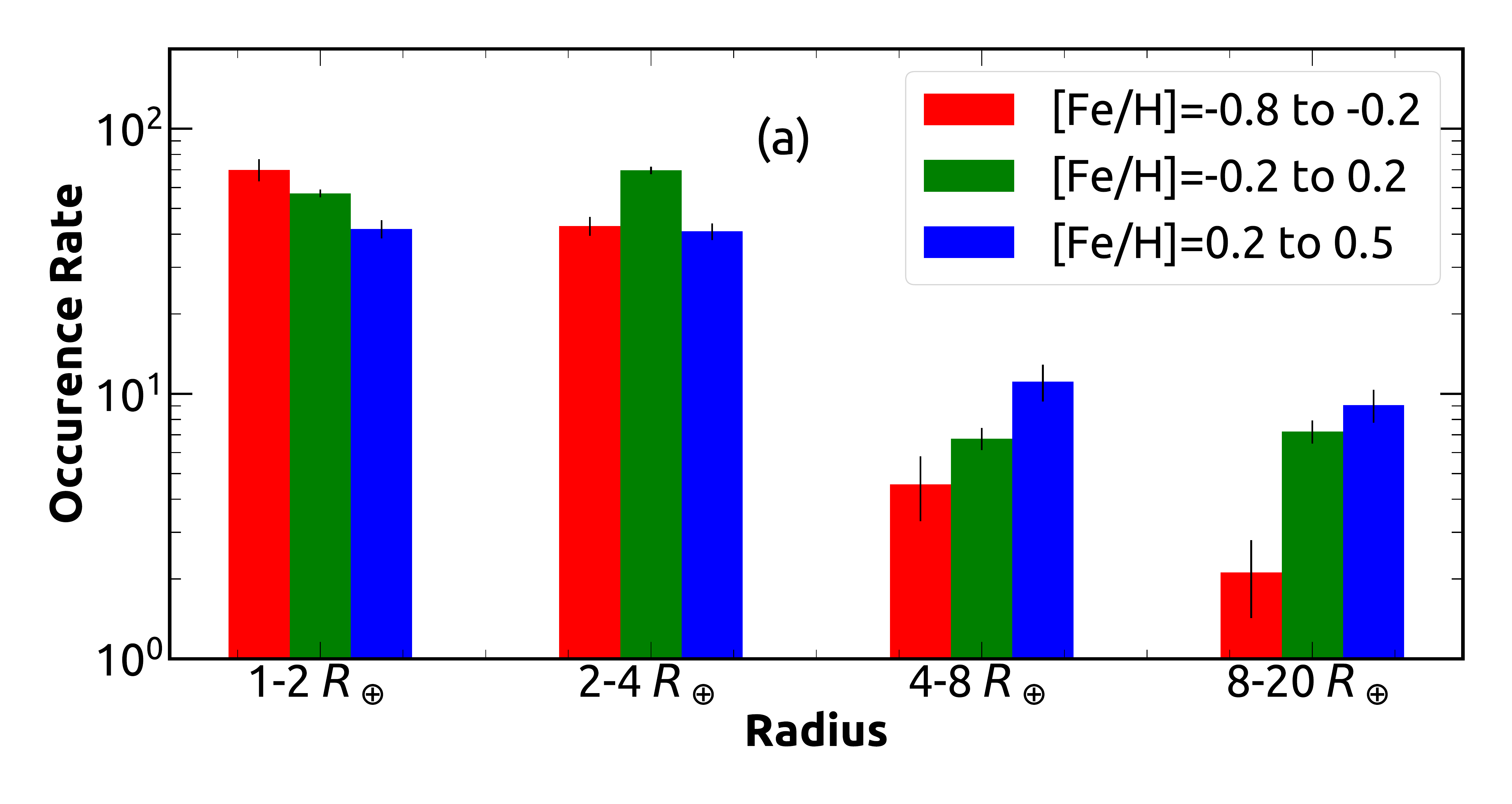}
\plotone{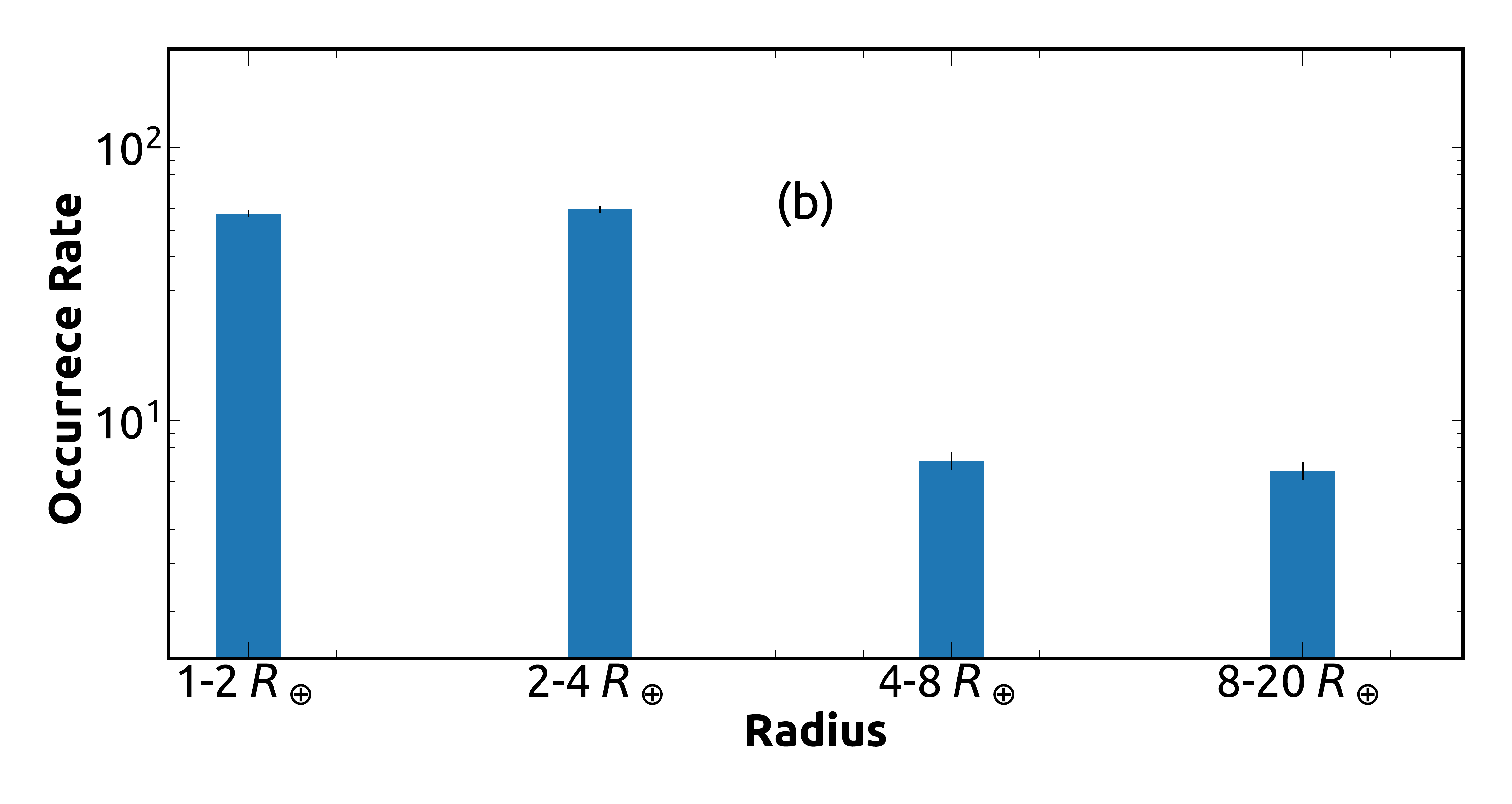}
\plotone{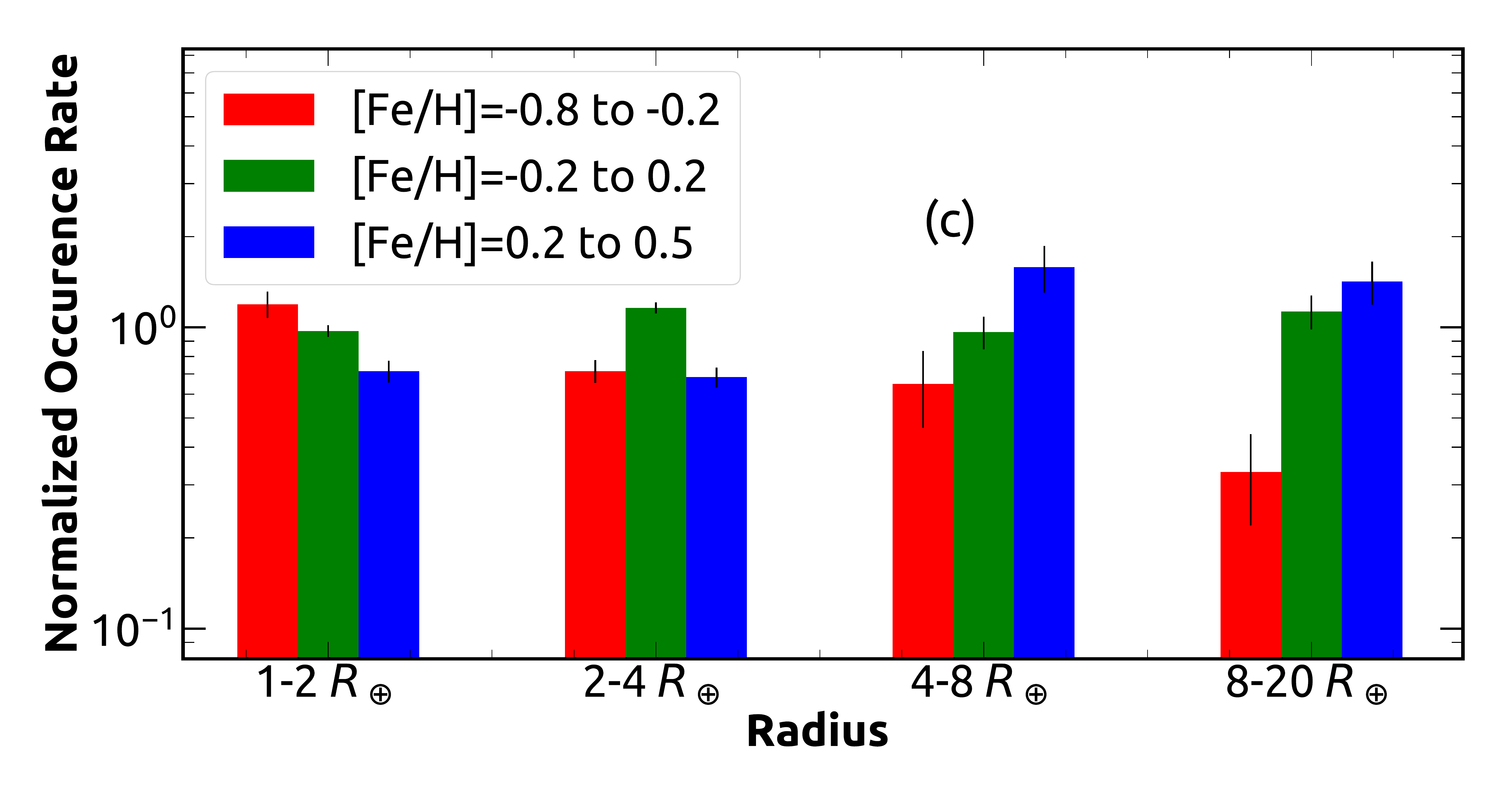}

\caption{(a) Occurrence rate of exoplanets as a function of planetary radius and host star metallicity. (b) The total occurrence rate of the sample without subdividing it into different metallicity bins. (c) Normalized occurrence rate of exoplanets as a function of planetary radius and host star metallicity. The error bars in these plots are the Poissonian errors based on the number of planets in each bin. }
\label{Figure9}

\end{center}
\end{figure}

\subsection{Occurrence rate of exoplanets as a function of planet radius and host star metallicity }

Using the above mathematical prescription we calculated the occurrence rate (per 100 stars) as a function of host star metallicity and planetary radius for the DR25 sample (Figure~\ref{Figure9}(a)). We also calculated the total occurrence rate as a function of planet radius only for the DR25 sample (Figure~\ref{Figure9}(b)). Figure \ref{Figure9}(a) is a function of both the planetary radius and host star metallicity. In Figure \ref{Figure9}(b), we have removed the dependence on host star metallicity and calculated the occurrence rate only as a function of planetary radius. Figure \ref{Figure9}(a) and Figure \ref{Figure9}(b) show that the occurrence rate is a much stronger function of the planetary radius than host star metallicity. 

To remove the underlying dependence of occurrence rate of planets on the planetary radius we normalize the occurrence rate in Figure \ref{Figure9}(a) with the total occurrence rate (Figure \ref{Figure9}(b)) to get the normalized occurrence rate in Figure \ref{Figure9}(c). The normalized occurrence rate should be only a function of \hs metallicity. 

In Figure \ref{Figure9}(c), the normalized occurrence rate of exoplanets is shown for various host star metallicities. This occurrence rate for super-solar metallicity host stars (blue bin) increases as a function of the planetary radius. A reverse trend is seen for the normalized occurrence rate of planets around sub-solar metallicity host stars (red bin), where the occurrence rate decreases as a function of the planetary radius. The normalized occurrence rate of planets around solar-type metallicity host stars (green bin) is mostly constant. Figure \ref{Figure9}(c) shows that the host star with super-solar metallicities have higher occurrence rate for giant planets than metal-poor (solar and sub-solar metallicities) host stars. These results are consistent with Figures \ref{Figure4}, \ref{Figure7}, \ref{Figure8}, and \ref{Figure9} that indicate that giant planets preferentially form around metal-rich stars. These results are also consistent with previous results in literature \citep[e.g.,][]{gonzalez97,santos01,santos04,fischer05,johnson07,buchhave12,mann13,buchhave14,buchhave15} and with those of \cite{CKS4}.

\section{HOST STAR METALLICITY AND PLANET’S ORBITAL PERIOD}
Previous studies on the orbital period of exoplanets and host star metallicity have revealed a dearth of planets with orbital periods less than 10 days around metal-poor stars \citep[e.g.,][]{beauge13,adibekyan13,mulders16,maldonado17b,wilson17, CKS4}. Host stars of planets with orbital periods greater than 10 days were more metal-poor than their short period counterparts. \cite{mulders16} and \cite{CKS4} showed that the occurrence rate of planets with periods less than 10 days was higher around metal-rich host stars than metal-poor host stars.

\begin{figure}
\begin{center}
\epsscale{1.25}

\plotone{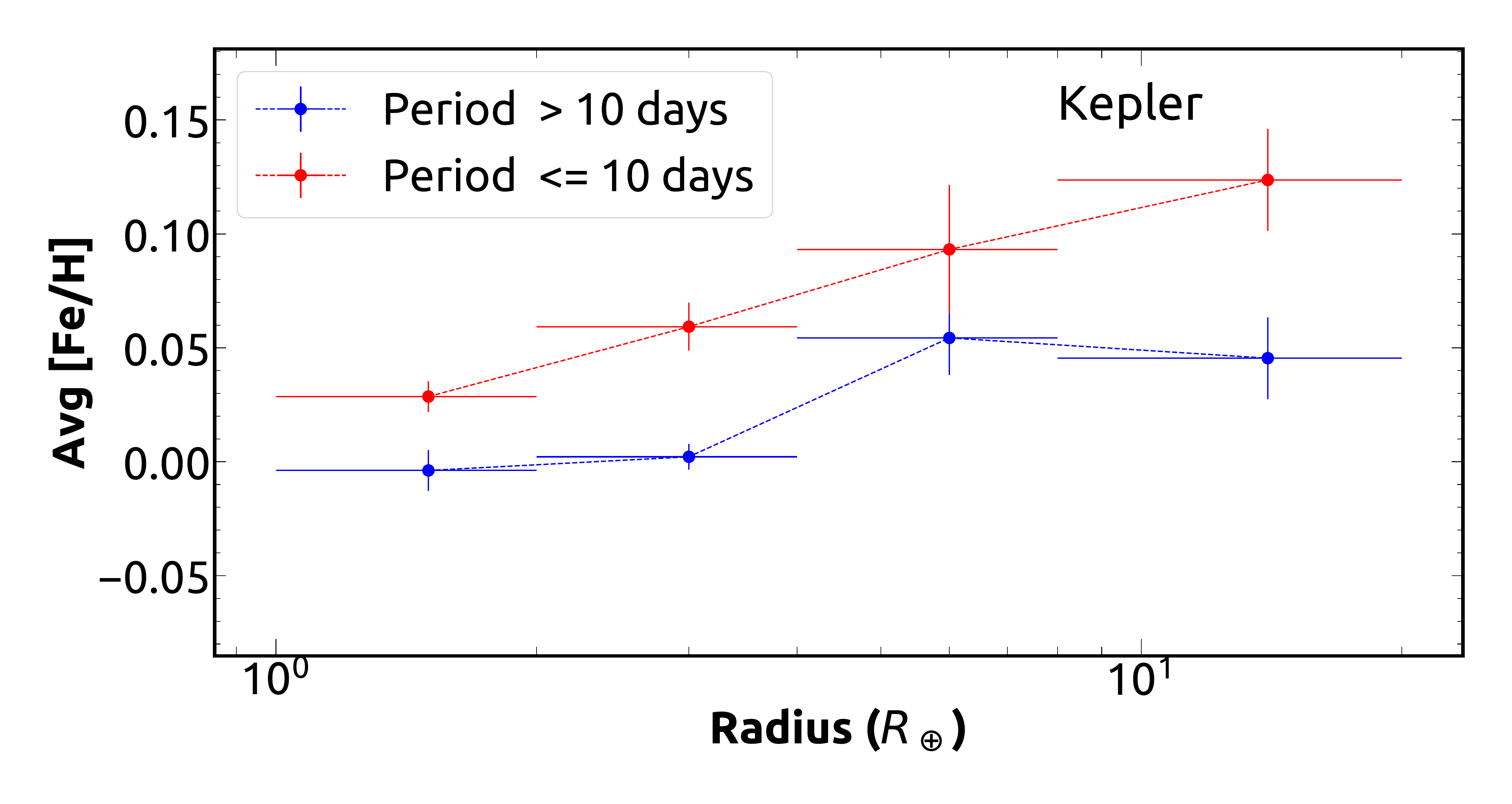}
\plotone{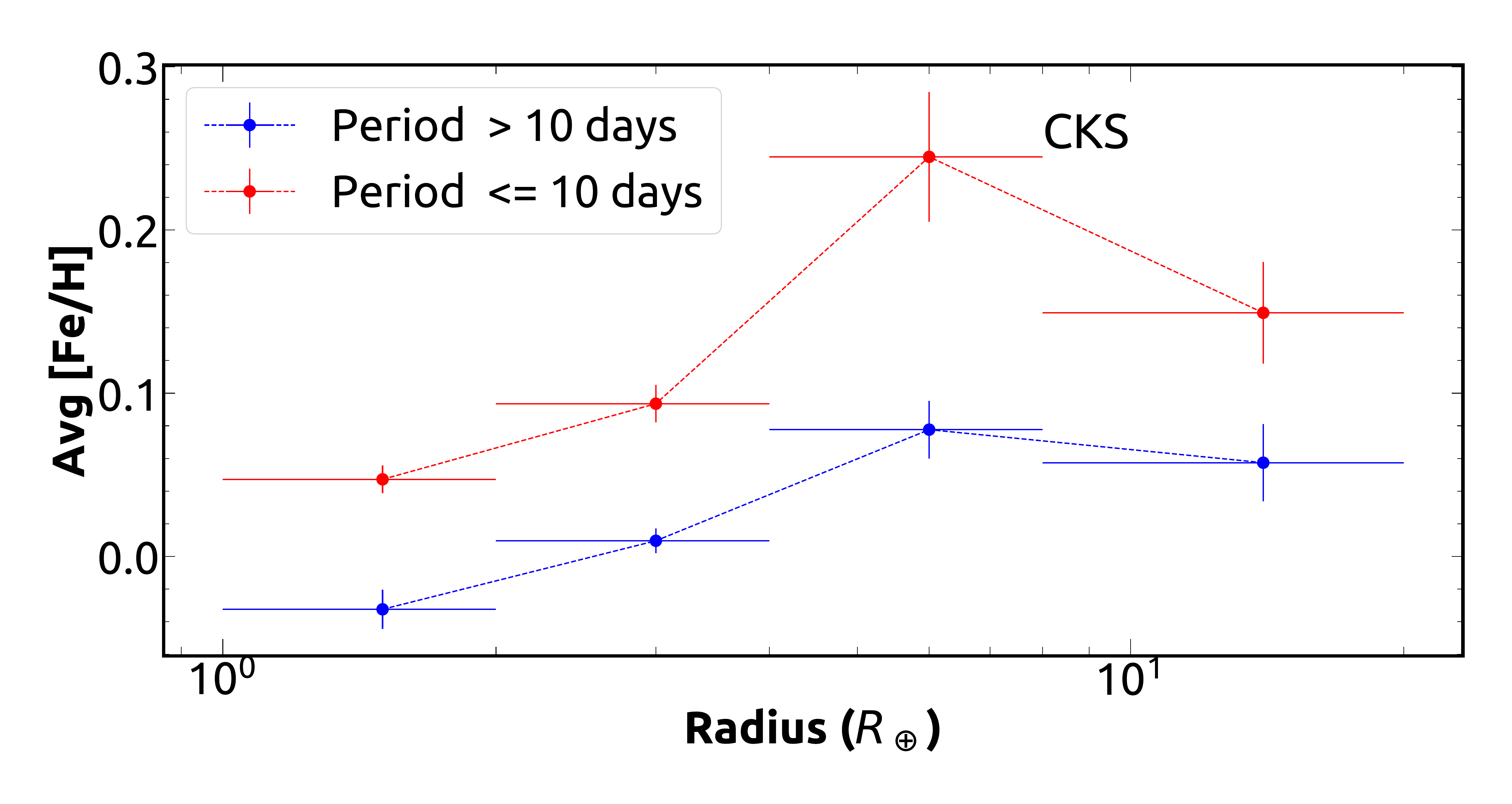}


\caption{Average host star metallicity as a function of the planetary radius for 2 different period bins. (a) Planetary radius and host star metallicity from the $Kepler$ DR25 sample, (b) planetary radius and host star metallicity from the CKS sample.  The error bars for X-axis is the width of the radius bins and for Y-axis we show the standard error in the mean of [Fe/H] as the error bar. }
\label{Figure10}

\end{center}
\end{figure}

\begin{figure}
\begin{center}

\epsscale{1.25}
\plotone{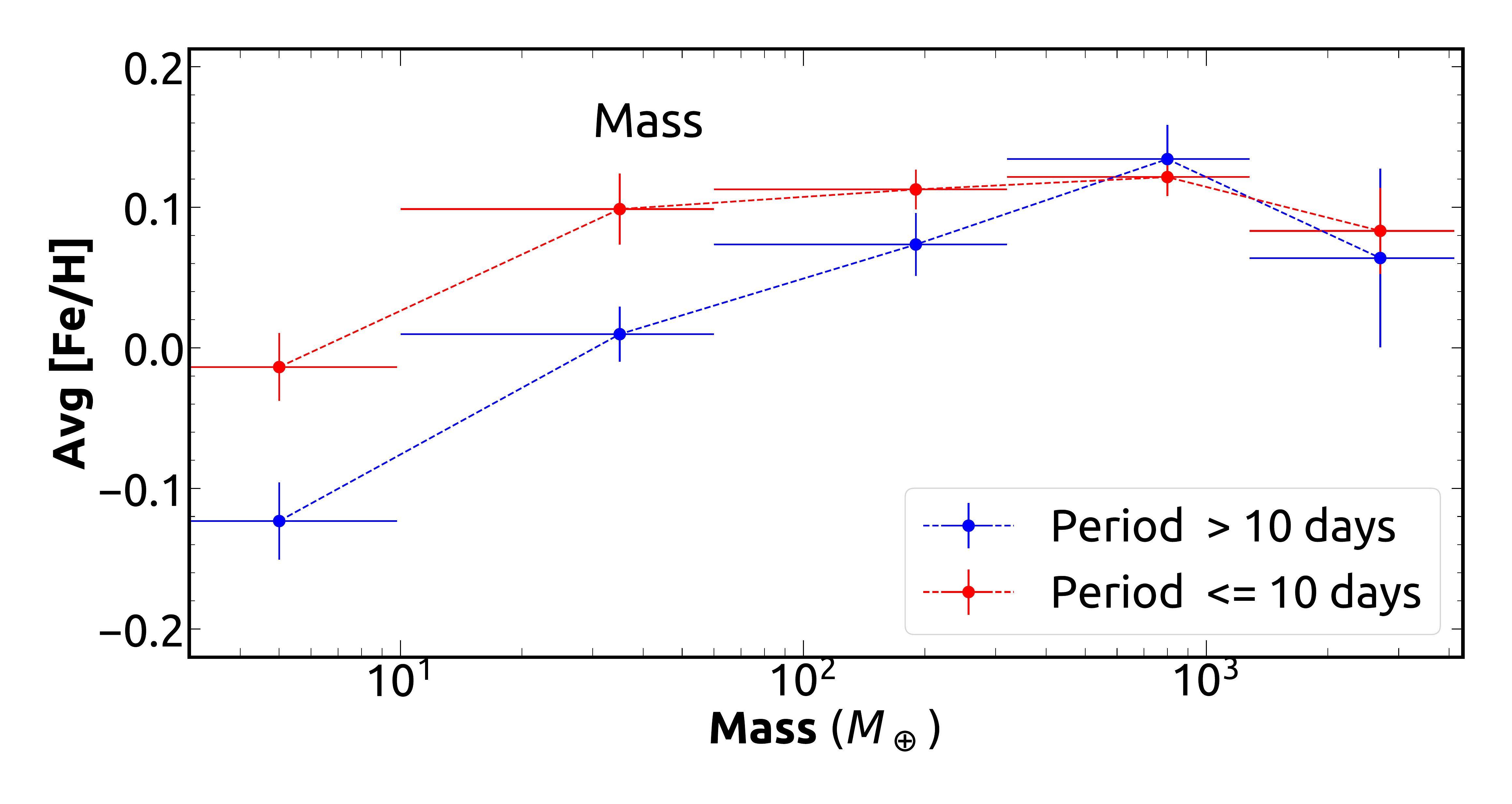}

\caption{Average host star metallicity as a function of the planetary mass. The planetary mass is from NASA Exoplanet Catalog and host star metallicity from SWEET-Cat catalog. The error bars for the X-axis is the width of the mass bins and for Y-axis we show the standard error in the mean of [Fe/H] as the error bar. }
\label{Figure11}

\end{center}
\end{figure}

\subsection{Average host star metallicity and orbital period}
In Figure \ref{Figure10}, we show the average host star metallicity as a function of planet radius for two bins of the orbital period. In Figure \ref{Figure10}(a) we divide the $Kepler$ planet candidate sample into two sub-groups: one having a period less than 10 days (red) and one having a period greater than 10 days (blue). Just like Figure \ref{Figure4} we bin the planet candidates into various radius bins following Section 3.1. In Figure \ref{Figure10}(b), we follow the same prescription for the CKS sample. From Figure \ref{Figure10} we find that for planets  with periods less than 10 days the host stars are usually more metal-rich than their longer period counterparts. 

In Figure \ref{Figure11}, we use the sample from Section 4 and show the average host star metallicity from SWEET-Cat as a function of the period for planets in various mass bins. We again divide the planet sample into two sub-groups: one having periods less than 10 days (red) and one having periods greater than 10 days (blue). We find that for planets with masses up to $ 50$\, $M_\oplus$ and periods less than 10 days the host stars are usually more metal richer than their longer period counterparts, but no such trend is found for giant planets.

\subsection{Occurrence rate of exoplanets as a function of orbital period and host star metallicity}

\begin{figure}
\begin{center}
\epsscale{1.25}

\plotone{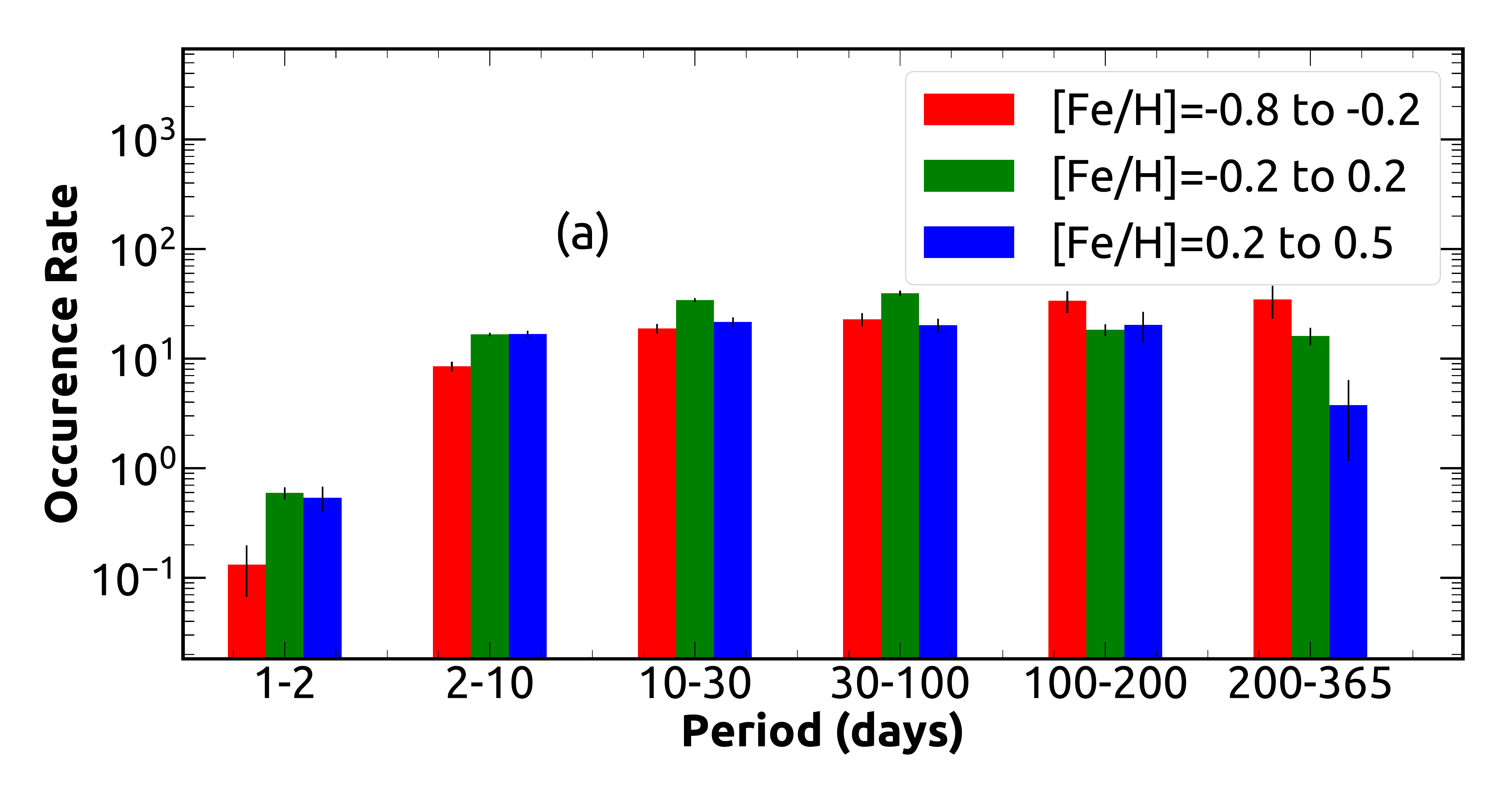}
\epsscale{1.33}
\plotone{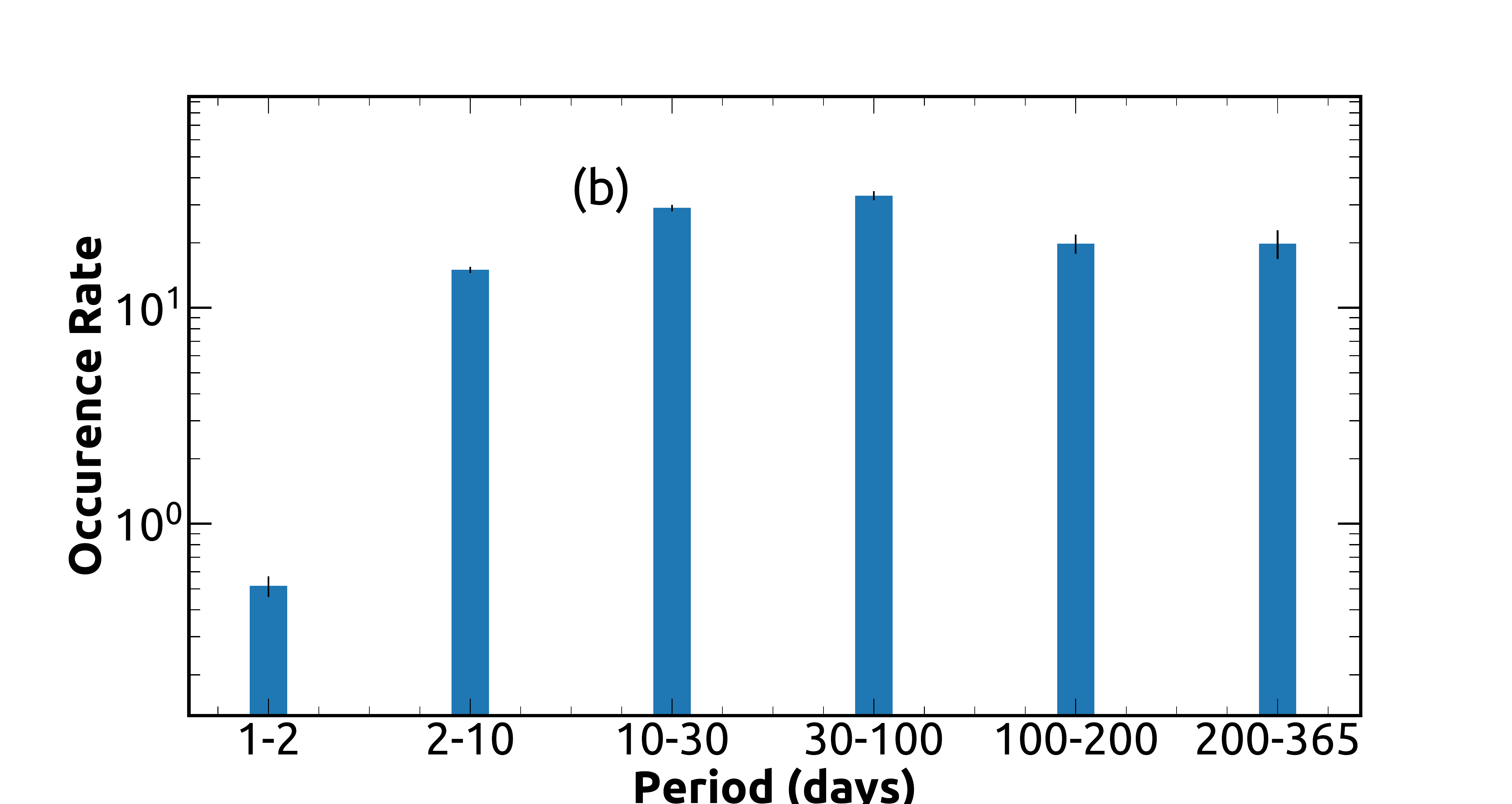}
\epsscale{1.25}
\plotone{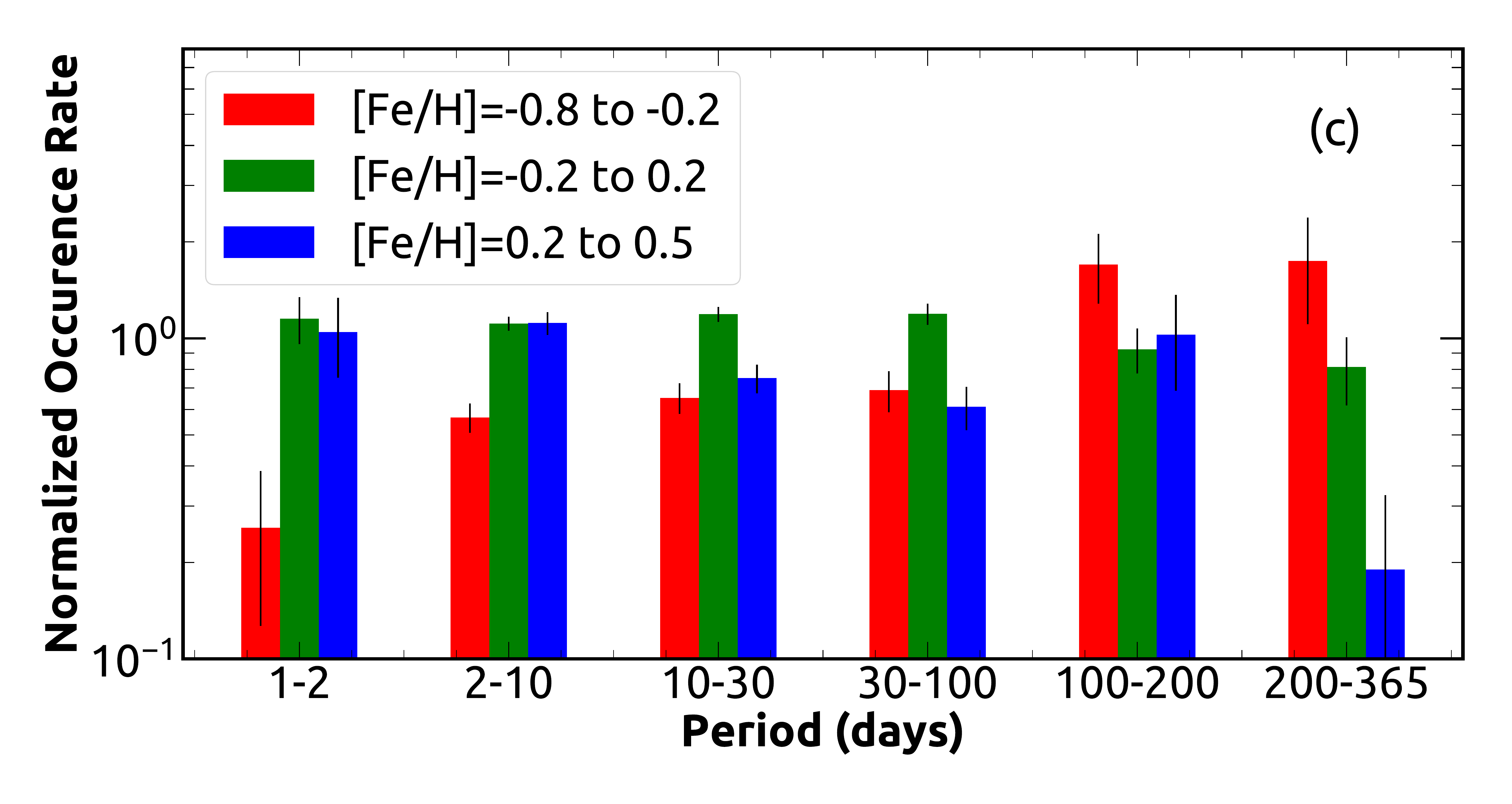}

\caption{ (a) Occurrence rate of exoplanets as a function of orbital period and host star metallicity for planets having radii less than $4\, R_\oplus$. (b) The total occurrence rate of the sample without subdividing it into different metallicity bins. (c) Normalized occurrence rate of exoplanets as a function of orbital period and host star metallicity. The error bars in these plots are the Poissonian errors from counting of planets. }
\label{Figure12}

\end{center}
\end{figure}

\begin{figure}
\begin{center}
\epsscale{1.25}

\plotone{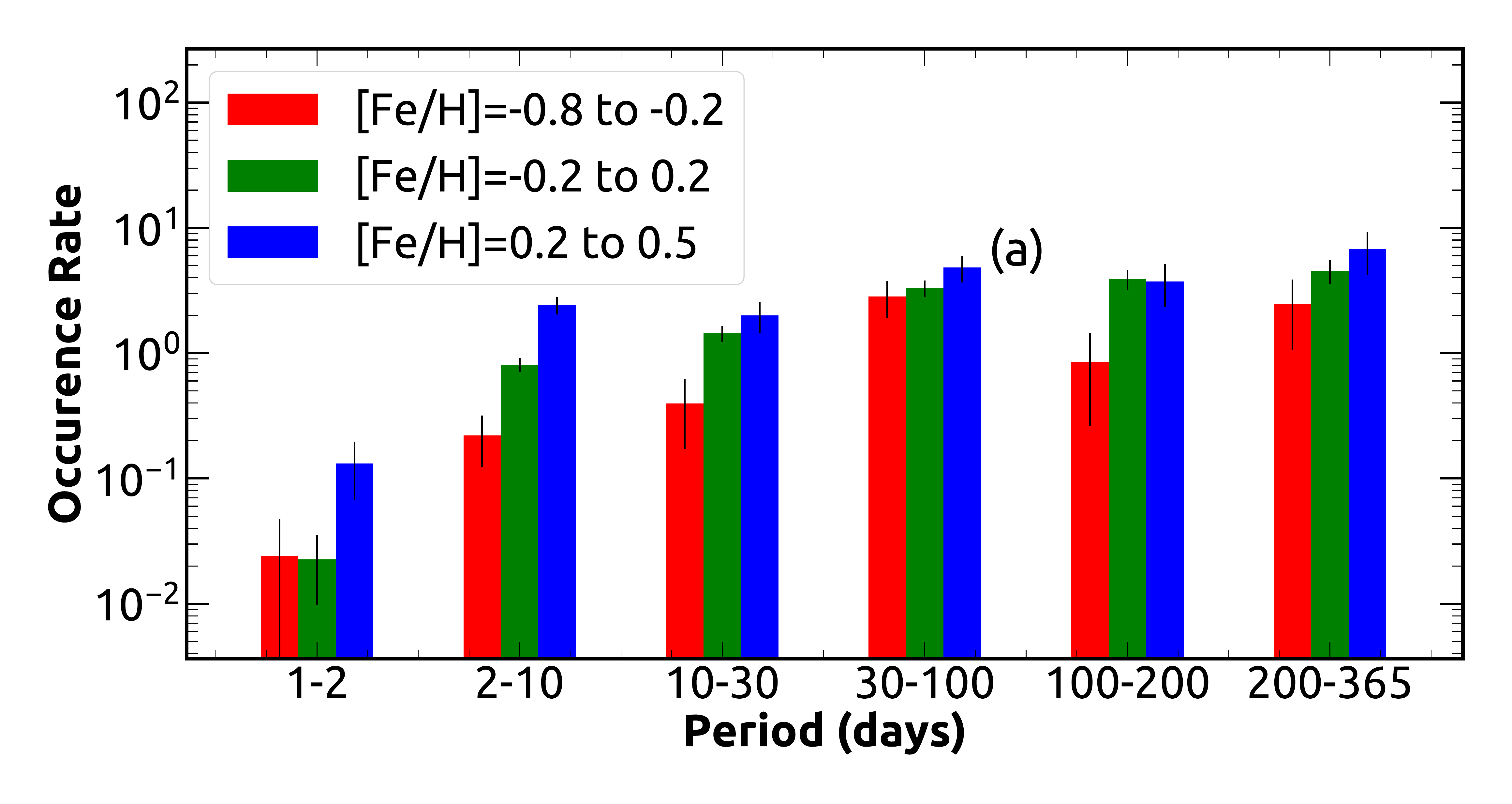}
\epsscale{1.17}
\plotone{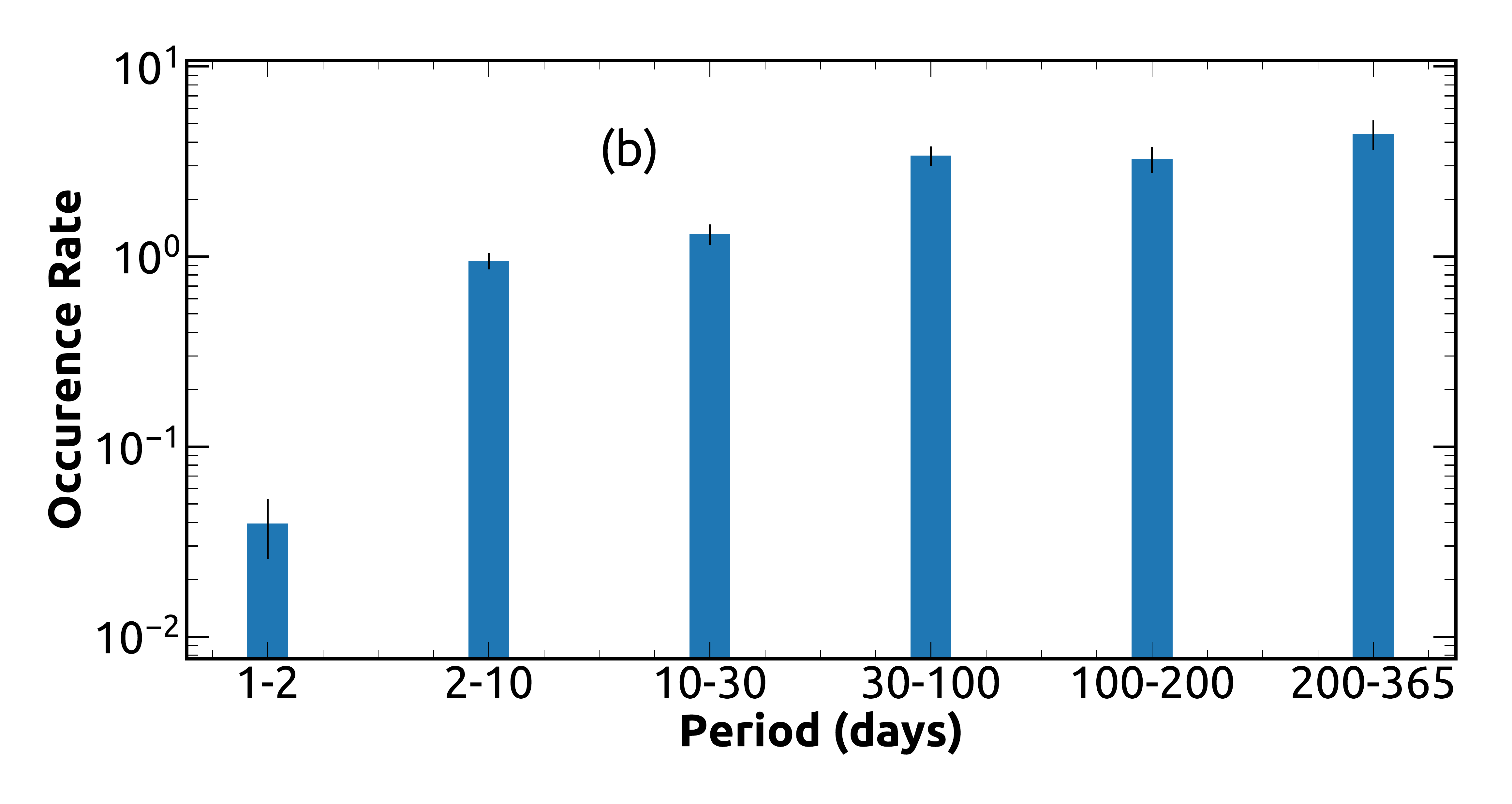}
\epsscale{1.25}
\plotone{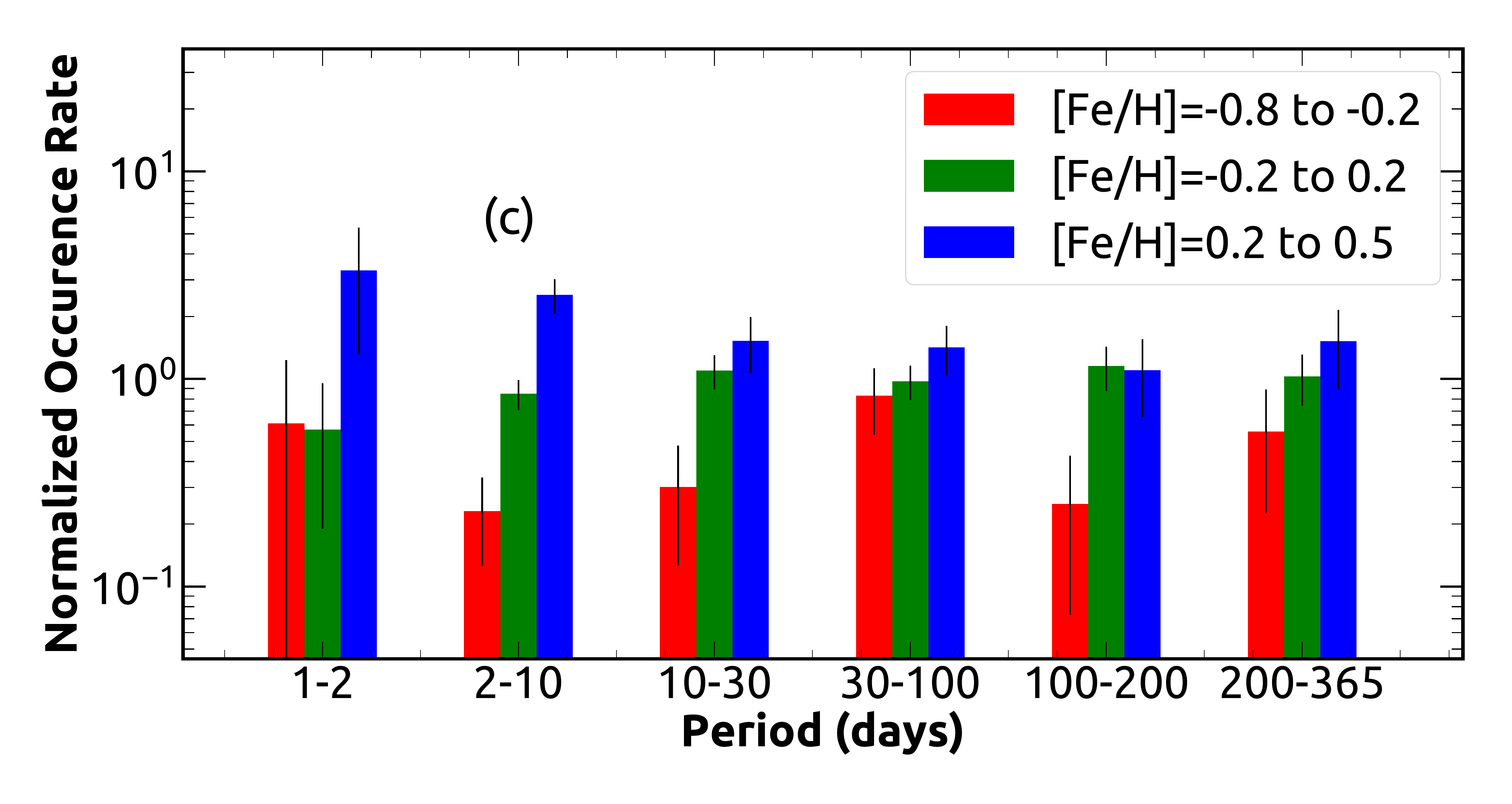}

\caption{ (a) Occurrence rate of exoplanets as a function of orbital period and host star metallicity for planets having radii between 4\Re and $20\, R_\oplus$. (b) The total occurrence rate of the sample without subdividing it into different metallicity bins. (c) Normalized occurrence rate of exoplanets as a function of orbital period and host star metallicity. The error bars in these plots are the Poissonian errors from counting of planets. }
\label{Figure13}

\end{center}
\end{figure}

In this section, we investigate the occurrence rate of exoplanets as a function of \hsm and orbital period for planets in two radius regimes: planets with $R_P \leq $ 4\Re (Figure \ref{Figure12}) and  planets with radius $R_P$ between $4 \, R_\oplus $ and $20~R_\oplus$ (Figure \ref{Figure13}). The occurrence rate as a function of host star metallicity and the orbital period of the planet is shown in sub-figures Figure \ref{Figure12}(a) and Figure \ref{Figure13}(a), the total occurrence rate as a function of orbital period is shown in sub-figures Figure \ref{Figure12}(b) and Figure \ref{Figure13}(b). From Figure \ref{Figure12}(a) and Figure \ref{Figure13}(a), we can infer that the occurrence rate is a much stronger function of the orbital period than the \hs metallicity. The occurrence rate as a function of period increases as the orbital period increases and plateaus around an orbital period of 10-30 days. To remove the underlying effect of orbital period on the occurrence rate, we normalize the occurrence rate in Figure \ref{Figure12}(a) and Figure \ref{Figure13}(a) with the total occurrence rate as a function of period (Figure \ref{Figure12}(b) and Figure \ref{Figure13}(b)) to calculate the normalized occurrence rate (Figure \ref{Figure12}(c) and Figure \ref{Figure13}(c)). 

Figure \ref{Figure12}(c) and Figure \ref{Figure13}(c) show that the normalized occurrence rate is highest for the super-solar host stars (blue bin) for planets with orbital periods less than 10 days. This indicates that the host star metallicity is higher for super-Earths, Neptunes, sub-Saturns, and Jupiters at orbital periods of less than 10 days. At orbital periods longer than 10-30 days in for planets with $R_P \leq 4 R_\oplus$  the normalized occurrence rates do not show any clear trend. 

The occurrence rates  (Figure \ref{Figure13}(a)) and the normalized occurrence rate  (Figure \ref{Figure13}(c)) of sub-Saturns and Jupiters are always highest for the super-solar (blue) bin. This shows that sub-Saturns and Jupiters are preferentially formed around metal-rich host stars. These results are consistent with the findings of Figure \ref{Figure10} and Figure \ref{Figure11}. Similar results have been derived by \cite{maldonado17b}, \cite{wilson17}, and \cite{CKS4}.

\section{DISCUSSION}

\subsection{Comparing our results from DR25 with \cite{CKS4} }
 
 Making use of the host star and planetary properties from the California Kepler Survey and the LAMOST spectra of the $Kepler$ field to derive properties of non-planet hosting stars, \cite{CKS4} estimated the occurrence rate of exoplanets as a function of host star metallicity, planetary radius, and orbital period.

They applied several filters to the sample including a magnitude limit for the host star, with a much smaller spectral range. They also excluded planets with a grazing transit and planets around host stars that have a nearby stellar companion. After applying these filters they have only 970 planets for which they compute the occurrence rate as a function of host metallicity, planet radius and orbital period. Whereas, we are using the latest $Kepler$ data release DR25 to calculate the occurrence rate as a function of host metallicity, planet radius, and orbital period.

On comparing our results with those of \cite{CKS4} we find similar trends between host star metallicity and planetary radius and orbital period though for a much larger sample of 2864 planets. Both studies show that the average \hsm increases as the planetary radius increases (see Figure \ref{Figure4} and \ref{Figure7}). Similarly, both studies show a higher occurrence rate for giants planet around metal-rich host stars confirming the early claims in the literature \citep[e.g.,][]{santos01,santos04,fischer05,johnson07,sozzetti09,johnson10,ghezzi10,sousa11,mortier12,buchhave12,mann13,buchhave14,buchhave15} (see Figure \ref{Figure10}). The average host star metallicity is also shown to be higher for planets with orbital periods of 10 days and less. The trends in the occurrence rate as a function of \hsm and period is similar in both studies (see Figure \ref{Figure12} and  Figure \ref{Figure13}). For small planets ($R_\mathrm{P} \leq 4\,R_\oplus$) with periods of about 10 days and less, \cite{CKS4} showed that the occurrence rate for metal-rich ([Fe/H]$>0$) host stars was higher than metal-poor ([Fe/H]$<0$) host stars. We also find similar results in Figure \ref{Figure12}, where we find a higher occurrence rate for the solar and super-solar metallicity bins for small planets ($R_\mathrm{P} \leq 4\, R_\oplus$) with periods less than 10 days.

The preceding discussion shows that although we use stellar parameter values from the DR25 stellar catalog \citep{mathur17}, the planet occurrence rates and their overall behavior as functions of planet radius, orbital period, and host star metallicity that we derive are consistent with those obtained by \cite{CKS4} using higher precision stellar parameters from high-resolution spectroscopy. 

Ongoing and upcoming surveys such as K2 and TESS will observe a significantly larger number of stars than $Kepler$. High-resolution spectroscopy follow-up to estimate stellar parameters for such a large sample well be likely to extremely resource intensive and time-consuming. Therefore, it is encouraging to know that stellar parameters that are less precise, such as the ones derived from broadband photometry (as was done for the KIC and which still constitute more than half of all the values of KOI host stars in the DR25 stellar table), can be useful to reveal trends between planetary and stellar parameters.

\subsection{Disk instability as the formation mechanism for super-Jupiter planets}
It is clear from the analysis so far that the metallicity of the host star is strongly correlated with the radius and mass of the planet. We find that as the radius of the planet increases, the \hsm increases. A similar trend is seen for planetary mass as well: as the mass of the planet increases, the average host star metallicity also increases; this trend, however, reverses at about $4\, M_\mathrm{J}$ after which, as the mass of the planet increases, the host star metallicity decreases. This trend of decreasing host star metallicity as the mass of the secondary increases extends even into the brown dwarf regime and low mass stars \citep{schlaufman18}.

The observed correlation of \hsm increasing with the increase in radius/mass of the planet is consistent with the predictions of the core accretion
model of planet formation where a 10 to 15\Me core needs to form before the planet can accrete gas and grow      \citep{mizuno80,pollack96,rice03,ida04,alibert04,laughlin04,ida05,kornet05,jhonson12,mordasini12}. Both the star and the protoplanetary disk (from which the planets form) form out of the same cloud material. A metal-rich star indicates that the cloud from which the star formed was also metal-rich. This means that the protoplanetary disk too would be metal-rich. Since a metal-rich disk means more solid material available to form the planetesimals, the core can form faster and grow before the disk dissipates.

Our results seem to indicate that  super-Jupiters ($M_\mathrm{P}> 4\,M_\mathrm{J}$) are preferentially found around metal-poor stars compared to stars that host Jupiters ($1-4 \,M_\mathrm{J}$) (see also \cite{santos17}). An explanation for this observed bimodal population of giant planets is that the Jupiters ($M_\mathrm{P}\leq  4\,M_\mathrm{J}$) are formed via core accretion whereas super-Jupiters ($M_\mathrm{P}> 4\,M_\mathrm{J}$) are formed via the disk instability or the gravitational instability \citep[e.g.,][]{boss97,rafikov05,stamatellos08,boley09,cai10,boss10,boss11}. The disk instability model assumes that the protoplanetary disk was massive enough to be unstable due to its own self-gravity. The major difference between this scenario and the core accretion model is the fact that in the disk instability model the solid components of the disk do not play a direct role in the process of planet formation. Therefore, the properties of planets formed via disk instability will not be as strongly correlated with host star metallicity as is the case for planets formed via core accretion model. Simulations of disk fragmentation have been shown to produce giant planets with a wide range in host star metallicity \citep{boss02}. 

We also find that the metallicity of host stars with brown dwarf companions is also lower compared to that of host stars of Jupiters  ($1-4 \, M_\mathrm{J}$). The fact that masses of super-Jupiters ($M_\mathrm{P}> 4\, M_\mathrm{J}$) and that of low mass brown dwarfs are comparable and that their host star metallicities are similar possibly suggest a common formation mechanism for both. Brown dwarfs in orbit around a more massive companion are thought to form via disk instability \citep[e.g.,][]{bate03,bate12, Kratter16}, lending further credence to the claim that super-Jupiters could form via disk instability. Similar results have recently been reported by \cite{santos17}, \cite{maldonado17} and \cite{schlaufman18} as well.

\subsection{Relationship between host star metallicity and planet orbital period}

Our analysis shows that planets smaller (and less massive) than Jupiter that have orbital periods less than 10 days are preferentially found around higher metallicity host stars compared to the ones with longer periods. A similar result was found by \cite{mulders16} and \cite{CKS4} but for a smaller sample. 

One of the suggested explanations for this is that the hot super-Earths and Neptunes are the remnants of gas giants whose atmospheres have been eroded due to the extreme environment \citep{lopez12,lundkvist16,mazeh16}. The other possible explanation is that these hot planets might have started off like hot Jupiters but were not able to accrete the gas quickly before the disk dissipated \citep{mulders16}.

Alternatively, this result can be explained by examining where the planets halt their migration \citep{mulders16}. Planet-forming disks around metal-rich stars will have more solid material in them. Such metal-rich disks can support planet migration much closer to the host star. This leads
to planets migrating and being found much closer in for
metal-rich stars \citep{mulders16,wilson17,CKS4}.

\section{Summary}

In this paper, we have studied the dependence of the observed properties (radius, mass, and orbital period) of exoplanets on their host star metallicity based on an analysis of more than 2800 $Kepler$ exoplanet candidates using the latest $Kepler$ data release DR25. Both the stellar and planetary properties were taken from the DR25 release except for stellar and planetary radii for which we used improved estimates based on $Gaia$ DR2 from \cite{berger18}. This is the largest sample for which such a study has been carried out so far. The results presented in this paper are consistent with all the previous work in literature \citep{gonzalez97,santos01,santos04,fischer05,johnson07,sozzetti09,johnson10,buchhave12,mann13,buchhave14,buchhave15,schlaufman15,wang15,santos17,mulders16,wilson17,maldonado17,schlaufman18,CKS4} though for a much larger sample of exoplanets. 

Most of the planet host stars in the DR25 sample have their properties determined from spectroscopy (about 60~\% of the total sample), whereas most of  the non-planet hosts have their properties, particularly, the metallicities derived from photometry. Although we have applied corrections in order make the spectroscopic and photometric metallicities consistent with each other, the metallicities derived from photometry have relatively larger uncertainties and are less accurate compared to spectroscopically determined metallicities. The occurrence rate calculated from such a mixed sample may not be as robust as that calculated with a pure spectroscopic sample for both host stars and non-host stars. But as stated earlier, the upcoming exoplanet surveys will be targeting several hundred thousands of stars; obtaining high-resolution spectra for all the host stars would be not possible.  Hence it is reassuring that even with a mixed sample of host star properties we are able to obtain results that have been derived previously using a much more homogeneous sample. 

Our main results are summarized below.

\begin{enumerate}
\item We investigated the correlation between the radius or mass of the planet and the host star metallicity. We find that the host star metallicity, on average, increases with increasing planet radius/mass up to about 10$\, R_\oplus$ or $1\, M_\mathrm{J}$. This is consistent with the predictions of the core accretion model of planet formation. In the case of planetary mass, we further show that as the mass of the planet increases above 4 \Mj the average host star metallicity decreases. This indicates that super-Jupiters possibly have a different formation mechanism than the Jupiter-size giant planets.

\item We calculated the occurrence rate of planets as a function of planetary radius and host star metallicity to account for known observational biases and selection effects. We find that the occurrence rate is a much stronger function of radius than host star metallicity. Therefore, in order to study the occurrence rate as a function of host star metallicity alone, we computed the normalized occurrence rate by removing the effect of planetary radius on the occurrence rate. We find that the normalized occurrence rate for super-solar metallicity host stars increases as a function of the planetary radius. This is consistent with the host star metallicity rising with increasing planet radius.

\item We further investigated the correlation between the host star metallicity and the orbital period of the planet. Planets with orbital periods less than 10 days appear to be more frequent around higher metallicity stars compared to planets with orbital periods longer than 10 days. For sub-Saturn and Jupiter-size planets ($R_\mathrm{P}$ between 4\Re and 20$\, R_\oplus$), the occurrence rate is the highest for host stars with super-solar metallicity ([Fe/H]$>$ 0.2) for all orbital periods. This further supports the idea that these planets are preferentially formed around metal-rich stars.

\end{enumerate}

\section{Acknowledgment }
We thank the referee for their insightful comments and suggestions that have led to the improvement of the final manuscript.

C.M. acknowledges the support from the Swiss National Science Foundation under grant BSSGI0$\_$155816 ``PlanetsInTime''. Parts of this work have been carried out within the frame of the National Center for Competence in Research PlanetS supported by the SNSF. This research has made use of the NASA Exoplanet Archive, which is operated by the California Institute of Technology, under contract with the National Aeronautics and Space Administration under the Exoplanet Exploration Program. This research has also made use of NASA’s Astrophysics Data System Abstract Service and of the SIMBAD database, operated at CDS, Strasbourg, France.


\end{document}